\documentclass[twocolum,showpacs,preprintnumbers,amsmath,amssymb]{revtex4}
\usepackage{graphicx}
\usepackage{dcolumn}
\usepackage{bm}

\begin{document}

\preprint{OU-TAP-236}
\preprint{YITP-04-47}

\title{Reconstructing the Primordial Spectrum 
with CMB Temperature and Polarization}

\author{Noriyuki Kogo$^{1,2}$}
 \email{kogo@yukawa.kyoto-u.ac.jp}
\author{Misao Sasaki$^2$}
 \email{misao@yukawa.kyoto-u.ac.jp}
\author{Jun'ichi Yokoyama$^1$}
 \email{yokoyama@vega.ess.sci.osaka-u.ac.jp}

\affiliation{$^1$Department of Earth and Space Science, 
       Graduate School of Science, Osaka University, 
       Toyonaka 560-0043, Japan}
\affiliation{$^2$Yukawa Institute for Theoretical Physics, 
       Kyoto University, Kyoto 606-8502, Japan}

\begin{abstract}
We develop a new method to reconstruct the power spectrum 
of primordial curvature perturbations, $P(k)$, 
by using both the temperature and polarization spectra 
of the cosmic microwave background (CMB). 
We test this method using several mock primordial spectra 
having non-trivial features including the one with an oscillatory component, 
and find that the spectrum can be reconstructed with a few percent accuracy 
by an iterative procedure in an ideal situation in which 
there is no observational error in the CMB data. 
In particular, 
although the previous ``cosmic inversion'' method, 
which used only the temperature fluctuations, 
suffered from large numerical errors 
around some specific values of $k$ that correspond to nodes 
in a transfer function, 
these errors are found to disappear almost completely in the new method. 
\end{abstract}

\maketitle

\section{Introduction} \label{INTRO}

The cosmic microwave background (CMB) anisotropies contain 
important pieces of information on physics of the early universe. 
The recent precise data of the Wilkinson Microwave Anisotropy Probe (WMAP) 
tell us that our universe is consistent with a spatially flat universe 
dominated by a cosmological constant ($\Lambda$) 
and cold dark matter (CDM), the so-called $\Lambda$CDM model, 
with slow-roll inflation at its early stage 
which has generated Gaussian, adiabatic, and nearly scale-invariant primordial 
fluctuations~\cite{WMAPBASIC,WMAPPARA,WMAPINF,WMAPGAUSS}. 

Nevertheless, it is reported that the observed CMB angular power spectrum 
may have some non-trivial features such as 
lack of power on large scales~\footnote{
This feature was already suggested by COBE observation~\cite{COBE,JF94} 
and possible explanation had been proposed~\cite{JY99}.}, 
running of the spectral index, 
and oscillatory behaviors of the spectrum on intermediate scales. 
To explain these non-trivial features, 
many authors focused on 
the primordial power spectrum of the curvature perturbation, $P(k)$, 
and proposed possible inflation 
models~\cite{BFZ03,CPKL03,CCL03,FZ03,TMB03,KT03,PFZ04,HM04,LMMR04,
KYY03,CST03,FLZZ03,YY03,HL03A,HL03B,HL03C,KS03,YY04,AN04,RE04,BFM03,DK03,
MR04A,MR04B,MR04C,KTT04,HS04}. 
However, equally important is to understand 
the implication of these observed possible non-trivial features 
on the primordial curvature spectrum without a theoretical prejudice. 
In this respect, there have been several attempts 
to reconstruct the primordial spectrum using the WMAP data 
by model-independent methods~\cite{MW03A,MW03B,BLWE03,SH04,SS03,TDS04}. 

In our previous work, 
we developed a method to reconstruct the primordial spectrum directly 
from the observed CMB anisotropy~\cite{MSY02,MSY03}, 
which we call the cosmic inversion method, 
and applied it to the WMAP first-year data~\cite{KMSY04}. 
Compared with other reconstruction methods such as 
the binning, wavelet band powers, and direct wavelet expansion method, 
we have shown that 
our method can reproduce fine features in $P(k)$ with a resolution of 
$\Delta k \simeq 3.7\times10^{-4}\,{\rm Mpc}^{-1}$ 
which roughly corresponds to $\Delta\ell \simeq 5$ 
in the angular power spectrum, $C_\ell$. 
Performing a statistical analysis of the reconstructed $P(k)$, 
we have found that there are some possible deviations 
from scale-invariance around 
$k \simeq 1.5\times10^{-2}\,{\rm Mpc}^{-1}$ 
and $2.6\times10^{-2}\,{\rm Mpc}^{-1}$~\cite{KMSY04}. 

The method we developed, however, used only 
the temperature-temperature (TT) angular power spectrum. 
The CMB anisotropy contains another important information, the polarization. 
During the recombination epoch the linear polarization of the CMB 
is generated by the Thomson scattering 
of the local quadrupole component of the temperature anisotropy~\cite{AK96}. 
From the perspective of a cross check, 
it is important to reconstruct $P(k)$ 
taking the CMB polarization into account. 
The CMB polarization was detected 
by the degree angular scale interferometer (DASI) 
for the first time~\cite{DASI} 
and more precise data 
of the temperature-polarization (TE) angular power spectrum 
were released by the WMAP recently~\cite{WMAPPOL}. 
In the future, much more precise observation of the polarization 
will be carried out by the Planck 
satellite~\footnote{http://www.rssd.esa.int/index.php?project=PLANCK}. 

In this paper, we present an improved cosmic inversion method 
that takes account of the CMB polarization spectrum, 
and test our new method for various shapes of $P(k)$. 
That is, we calculate the CMB temperature and polarization spectra 
for a given $P(k)$ and reconstruct it from thus obtained CMB data. 
As a first step, we neglect observational errors in numerical tests 
and assume that the CMB temperature and polarization spectra 
are completely known. 
We calculate the CMB power spectra based on 
CMBFAST~\footnote{http://www.cmbfast.org/} 
but we adopt a much finer resolution than the original one 
in both $k$ and $\ell$. 
Our inversion method is an iterative method based on 
approximate formulas for CMB temperature and polarization spectra 
and correction factors that adjust the errors in the approximate formulas. 

This paper is organized as follows. 
In Sec.~\ref{BASIC}, we review the basic theory of the CMB polarization. 
In Sec.~\ref{METHOD}, we extend our cosmic inversion method 
and propose a new inversion formula that 
uses both the CMB temperature and polarization spectra. 
In Sec.~\ref{TESTS}, we test our new method. 
We find that the primordial spectrum can be reconstructed 
with much better accuracy. 
In Sec.~\ref{PARAMETERS}, we propose the idea of 
constraining the cosmological parameters based on our new method. 
Finally, Sec.~\ref{CONCLUSION} is devoted to conclusion. 

\section{Basic Theory} \label{BASIC}

Here we summarize basic equations 
for the CMB temperature fluctuations and polarization 
based on~\cite{ZS97,KKS97,HW97}, 
and their approximate formulas which are used in our inversion method. 
We deal only with scalar-type perturbations 
and assume a spatially flat universe 
with Gaussian and adiabatic primordial fluctuations. 
We leave inclusion of tensor-type perturbations for future work.

It is convenient to describe polarized radiation by 
the Stokes parameters~\cite{SC60,RL60}. 
Consider a monochromatic electromagnetic wave 
propagating in the $z$-direction with an angular frequency $\omega_0$. 
The components of the electric field are written as 
\begin{eqnarray}
E_x &=& {\cal E}_x(t) \cos [\omega_0 t-\varphi_x(t)],
\label{EX} \\
E_y &=& {\cal E}_y(t) \cos [\omega_0 t-\varphi_y(t)].
\label{EY}
\end{eqnarray}
The Stokes parameters are defined as
\begin{eqnarray}
I &\equiv& \bigl\langle {\cal E}_x{}^2+{\cal E}_y{}^2 \bigl\rangle,
\label{STOKESI} \\
Q &\equiv& \bigl\langle {\cal E}_x{}^2-{\cal E}_y{}^2 \bigl\rangle,
\label{STOKESQ} \\
U &\equiv& \bigl\langle 2{\cal E}_x {\cal E}_y
           \cos(\varphi_x-\varphi_y) \bigl\rangle,
\label{STOKESU} \\
V &\equiv& \bigl\langle 2{\cal E}_x {\cal E}_y
           \sin(\varphi_x-\varphi_y) \bigl\rangle,
\label{STOKESV}
\end{eqnarray}
where the brackets denote the time average 
over the modulation of the amplitudes and the phases. 
Since the Thomson scattering generates only linear polarization, 
we can neglect $V$ which describes the degree of circular polarization. 
Under a rotation of an angle $\psi$ on the $(x,y)$-plane, 
$I$ is invariant while $Q$ and $U$ are transformed as 
\begin{eqnarray}
Q' &=&  Q\cos 2\psi+U\sin 2\psi,
\label{QTRANS} \\
U' &=& -Q\sin 2\psi+U\cos 2\psi,
\label{UTRANS}
\end{eqnarray}
which are identical to 
\begin{eqnarray}
(Q \pm iU)'=e^{\mp2i\psi}(Q \pm iU).
\label{QUTRANS}
\end{eqnarray}
This means that $(Q \pm iU)$ have spins $\pm 2$, respectively. 

We denote the temperature fluctuations and the polarization of radiation 
at the conformal time $\eta$ and the comoving spatial position $\bm{x}$ 
propagating in the direction $\hat{\bm{n}}$ 
as $\Theta(\eta,\bm{x},\hat{\bm{n}})$ 
and $(Q \pm iU)(\eta,\bm{x},\hat{\bm{n}})$, respectively. 
In general, these are expanded using the normal modes defined as 
\begin{eqnarray}
_sG_{\ell m} \equiv
(-i)^\ell \sqrt{\frac{4\pi}{2\ell+1}}\, _sY_{\ell m}(\hat{\bm{n}}) \,
e^{i\bm{k}\cdot\bm{x}},
\label{MODE}
\end{eqnarray}
where $\bm{k}$ is the comoving wavenumber 
and $_sY_{\ell m}$ is the spin-$s$ weighted spherical harmonic 
function~\cite{NP66,GEA67,KST80}. 
For each Fourier mode 
we can work in the coordinate system where $\bm{k} \parallel \hat{\bm{z}}$. 
In the case of scalar perturbations, $m=0$, 
\begin{eqnarray}
\Theta(\eta,\bm{x},\hat{\bm{n}})
&=& \int \! \frac{d^3 \bm{k}}{(2\pi)^3} \,
    \sum_{\ell=0}^{\infty} \Theta_\ell(\eta,k) \, _0G_{\ell 0}
\nonumber \\
&=& \int \! \frac{d^3 \bm{k}}{(2\pi)^3} \, e^{i\bm{k}\cdot\bm{x}}
    \sum_{\ell=0}^{\infty} (-i)^\ell \Theta_\ell(\eta,k) P_\ell(\mu),
\label{MULTIT} \\
(Q \pm iU)(\eta,\bm{x},\hat{\bm{n}})
&=& \int \! \frac{d^3 \bm{k}}{(2\pi)^3} \,
    \sum_{\ell=0}^{\infty} (E_\ell \pm iB_\ell)(\eta,k) \,
    _{\pm 2}G_{\ell 0}
\nonumber \\
&=& \int \! \frac{d^3 \bm{k}}{(2\pi)^3} \, e^{i\bm{k}\cdot\bm{x}}
    \sum_{\ell=0}^{\infty} (-i)^\ell \sqrt{\frac{(\ell-2)!}{(\ell+2)!}}\,
    (E_\ell \pm iB_\ell)(\eta,k) P_\ell^2(\mu),
\label{MULTIP}
\end{eqnarray}
where $\mu \equiv \hat{\bm{k}} \cdot \hat{\bm{n}}$ 
and $P_\ell^2$ is the associated Legendre function, $P_\ell^m(\mu)$ with
$m=2$.  Here 
$E_\ell$ and $B_\ell$ represent the so-called E-mode and B-mode polarizations 
with parities  $(-1)^\ell$ and $(-1)^{\ell+1}$, respectively. 
Note that for scalar-type perturbations, 
$B_\ell$ vanishes because of its parity. 
Using these multipole moments, 
the angular power spectrum is expressed as 
\begin{eqnarray}
\frac{2\ell+1}{4\pi} C^{X\bar{X}}_\ell
=\frac{1}{2\pi^2} \int_0^\infty \! \frac{dk}{k} \,
 \frac{k^3 \bigl\langle X_\ell^*(\eta_0,k)
 \bar{X}_\ell(\eta_0,k) \bigl\rangle}{2\ell+1},
\label{CL}
\end{eqnarray}
where $X$ and $\bar{X}$ are either $\Theta$ ($=T$) or $E$. 
The Boltzmann equations for each $\Theta_\ell(\eta,k)$ and $E_\ell(\eta,k)$ 
can be transformed into the following integral form ($\ell \ge 2$): 
\begin{eqnarray}
\frac{\Theta_\ell(\eta_0,k)}{2\ell+1}
&=& \int_0^{\eta_0} \! d\eta \, \bigg\{
    \left[ (\Theta_0+\Psi) {\cal V}(\eta)
    +(\dot{\Psi}-\dot{\Phi}) e^{-\tau(\eta)}
    \right] j_\ell(k\Delta\eta) \nonumber \\
& & {}+V_b {\cal V}(\eta) j'_\ell(k\Delta\eta)
    +\frac{1}{2} \Pi_2 {\cal V}(\eta)
    \left[ 3j''_\ell(k\Delta\eta)+j_\ell(k\Delta\eta) \right]
    \bigg\},
\label{TL} \\
\frac{E_\ell(\eta_0,k)}{2\ell+1}
&=& -\frac{3}{2} \sqrt{\frac{(\ell+2)!}{(\ell-2)!}}
    \int_0^{\eta_0} \! d\eta \,
    \Pi_2 {\cal V}(\eta)\,
    \frac{j_\ell(k\Delta\eta)}{(k\Delta\eta)^2},
\label{EL}
\end{eqnarray}
where $\Pi_2 \equiv (\Theta_2-\sqrt{6} E_2)/10$, 
$\Delta\eta \equiv \eta_0-\eta$ 
with $\eta_0$ being the conformal time at present, 
and the overdot denotes a derivative with respect to the conformal time. 
Here $V_b$ is the baryon fluid velocity, 
$\Psi$ and $\Phi$ are the Newton potential 
and the spatial curvature perturbation in the Newton gauge, 
respectively~\cite{KS84}, and 
\begin{eqnarray}
{\cal V}(\eta) \equiv \dot{\tau} e^{-\tau(\eta)}, \quad
    \tau(\eta) \equiv \int_{\eta}^{\eta_0} \! \dot{\tau} d\eta,
\label{VIS&TAU}
\end{eqnarray}
are the visibility function and the optical depth for Thomson scattering, 
respectively. 
In the limit that 
the thickness of the last scattering surface (LSS) is negligible, we have 
${\cal V}(\eta) \approx \delta(\eta-\eta_*)$ and
$e^{-\tau(\eta)} \approx \theta(\eta-\eta_*)$, 
where $\eta_*$ is the recombination time 
when the visibility function is maximum~\cite{HS95}. 

\begin{figure}
\begin{center}
\includegraphics[width=10cm]{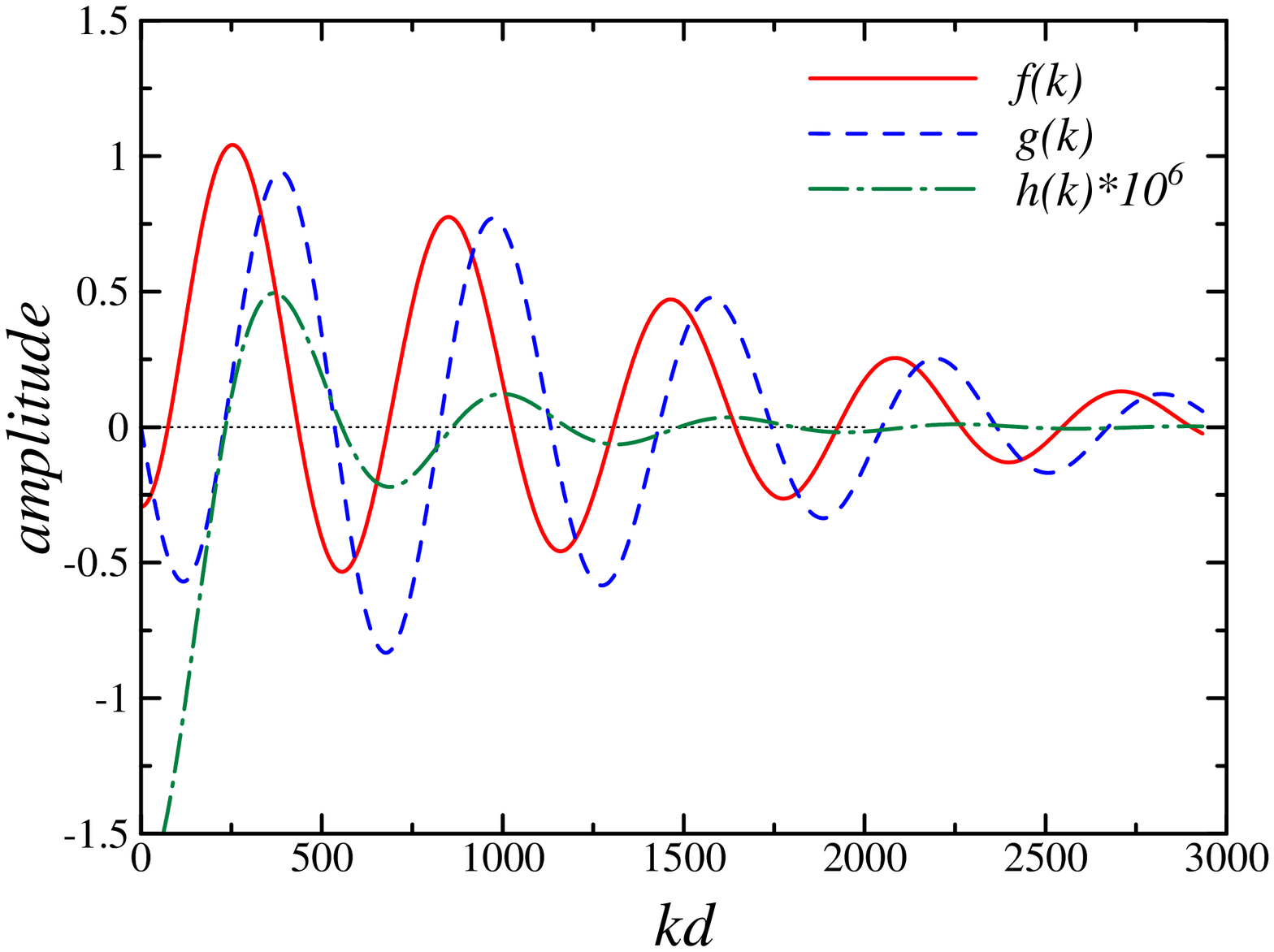}
\caption{Transfer functions $f(k)$ ({\it solid curve}), 
$g(k)$ ({\it dashed curve}), and $h(k)$ ({\it dash-dotted curve}) 
for $h=0.70$, $\Omega_b=0.050$, $\Omega_\Lambda=0.70$, 
$\Omega_m=0.30$ and $\tau=0.20$. 
The horizontal axis is the comoving wavenumber normalized by $d^{-1}$, 
where $d \simeq 1.36\times10^4\,{\rm Mpc}$ 
is the conformal distance from the present to the LSS. 
$f(k)$, $g(k)$ and $h(k)$ correspond to 
the monopole, dipole and quadrupole components, respectively, 
of the temperature anisotropy at the LSS. 
Since the amplitude of $h(k)$ is much smaller than those of $f(k)$ and $g(k)$, 
the amplitude of $h(k)$ is magnified by $10^6$. 
\label{TF}}
\end{center}
\end{figure}

As we did in our previous papers~\cite{MSY03,KMSY04}, 
we now take into account the thickness of the LSS approximately 
to perform the integrals in Eq.~(\ref{EL}) as well as 
in Eq.~(\ref{TL}) analytically.
For the polarization, this is crucial 
since the CMB polarization is mainly generated 
within the thickness of the LSS. 
The approximation is to neglect the oscillations of 
the spherical Bessel functions in the integrals. 
Applying this approximation to Eq.~(\ref{TL}) and (\ref{EL}), we have 
\begin{eqnarray}
\frac{\Theta_\ell(\eta_0,k)}{2\ell+1} &\approx&
 \left\{ \int_{\eta_{*{\rm start}}}^{\eta_{*{\rm end}}} \! d\eta \,
         \left[ (\Theta_0+\Psi) {\cal V}(\eta)
         +(\dot{\Psi}-\dot{\Phi}) e^{-\tau(\eta)} \right] \right\} j_\ell(kd)
         \nonumber \\ & &{}
+\left\{ \int_{\eta_{*{\rm start}}}^{\eta_{*{\rm end}}} \! d\eta \,
         \Theta_1(\eta,k) {\cal V}(\eta) \right\} j'_\ell(kd)
\,\,\equiv\,\,\frac{\Theta^{\rm app}_\ell(\eta_0,k)}{2\ell+1},
\label{TLAPP} \\
\frac{E_\ell(\eta_0,k)}{2\ell+1} &\approx&
 \sqrt{\frac{(\ell+2)!}{(\ell-2)!}}
 \left\{ -\frac{3}{2} 
 \int_{\eta_{*{\rm start}}}^{\eta_{*{\rm end}}} \! d\eta \,
 \frac{\Pi_2}{(k\Delta\eta)^2} {\cal V}(\eta) \right\} j_\ell(kd)
\,\,\equiv\,\,\frac{E^{\rm app}_\ell(\eta_0,k)}{2\ell+1},
\label{ELAPP}
\end{eqnarray}
where $d \equiv \eta_0-\eta_*$ is the conformal distance 
from the present to the LSS 
and $\eta_{*{\rm start}}$ and $\eta_{*{\rm end}}$ are the times 
when the recombination starts and ends, respectively. 
We have also replaced $V_b$ by $\Theta_1$ 
and neglected the quadrupole term in Eq.~(\ref{TL}), 
by adopting the tight coupling approximation~\cite{HS95}. 
The transfer functions, $f(k)$, $g(k)$, and $h(k)$ are defined by 
\begin{eqnarray}
\int_{\eta_{*{\rm start}}}^{\eta_{*{\rm end}}} \! d\eta \,
\left[ (\Theta_0+\Psi)(\eta,k){\cal V}(\eta)
      +(\dot{\Psi}-\dot{\Phi})(\eta,k) e^{-\tau(\eta)} \right]
&\equiv& f(k) \Phi(0,\bm{k}),
\label{TRANSF} \\
\int_{\eta_{*{\rm start}}}^{\eta_{*{\rm end}}} \! d\eta \,
\Theta_1(\eta,k){\cal V}(\eta)
&\equiv& g(k) \Phi(0,\bm{k}),
\label{TRANSG} \\
-\frac{3}{2} \int_{\eta_{*{\rm start}}}^{\eta_{*{\rm end}}} \! d\eta \,
 \frac{\Pi_2(\eta,k)}{(k(\eta_0-\eta))^2} {\cal V}(\eta)
&\equiv& h(k) \Phi(0,\bm{k}).
\label{TRANSH}
\end{eqnarray}
Given the cosmological parameters, 
one can calculate transfer functions numerically, e.g., by using CMBFAST. 
The results are depicted in Fig.~\ref{TF}. 

The primordial power spectrum of the curvature perturbation is defined as 
\begin{eqnarray}
P(k) \equiv \langle |\Phi(0,\bm{k})|^2 \rangle.
\label{PK}
\end{eqnarray}
Substituting Eqs.~(\ref{TLAPP}) and (\ref{ELAPP}) into Eq.~(\ref{CL}), 
we obtain the approximated TT, EE, and TE angular power spectra, 
\begin{eqnarray}
\frac{2\ell+1}{4\pi} C^{TT,\,{\rm app}}_\ell
&=& \frac{2\ell+1}{2\pi^2} \int_0^\infty \! \frac{dk}{k} \, k^3 P(k)
     \left[ f(k) j_\ell(kd)+g(k) j'_\ell(kd) \right]^2,
\label{CLTTAPP} \\
\frac{2\ell+1}{4\pi} C^{EE,\,{\rm app}}_\ell
&=& \frac{2\ell+1}{2\pi^2} \,\frac{(\ell+2)!}{(\ell-2)!} \,
 \int_0^\infty \! \frac{dk}{k} \, k^3 P(k) 
    \left[ h(k) j_\ell(kd) \right]^2,
\label{CLEEAPP} \\
\frac{2\ell+1}{4\pi} C^{TE,\,{\rm app}}_\ell
&=& \frac{2\ell+1}{2\pi^2}\sqrt{\frac{(\ell+2)!}{(\ell-2)!}}
 \int_0^\infty \! \frac{dk}{k} \, k^3 P(k) 
\left[ f(k) j_\ell(kd)+g(k) j'_\ell(kd) \right] h(k) j_\ell(kd).
\label{CLTEAPP}
\end{eqnarray}

\section{Inversion Method} \label{METHOD}

Here we describe our method of the reconstruction of $P(k)$. 
Let us first briefly review the case of the temperature fluctuations 
proposed in~\cite{MSY02,MSY03}. 
The CMB temperature fluctuations can be quantified 
by the angular correlation function defined as 
\begin{eqnarray}
C^{TT}(\theta)
\equiv \left\langle \Theta(\hat{\bm{n}}_1)
       \Theta(\hat{\bm{n}}_2) \right\rangle
     = \sum_{\ell=0}^{\infty} \frac{2\ell+1}{4\pi}
       C^{TT}_\ell P_\ell(\cos\theta),
\quad \cos\theta=\hat{\bm{n}}_1 \cdot \hat{\bm{n}}_2 \, .
\label{CRTT}
\end{eqnarray}
Here we introduce a new variable $r$ instead of $\theta$ defined as 
\begin{eqnarray}
r=2d \sin \frac{\theta}{2},
\label{CRAPP}
\end{eqnarray}
which is the conformal distance between two points on the LSS. 
Substituting Eq.~(\ref{CLTTAPP}) into Eq.~(\ref{CRTT}) 
and using the formula: 
\begin{eqnarray}
\sum_{\ell=0}^{\infty} (2\ell+1) P_\ell(\cos\theta) j_\ell{}^2(kd)
=\frac{\sin kr}{kr},
\label{BESSEL}
\end{eqnarray}
we can derive formulas for the approximated angular correlation functions 
in terms of $P(k)$. 
In the small-scale limit $r \ll d$, 
its derivative is written as 
\begin{eqnarray}
\frac{1}{r} \frac{\partial}{\partial r} 
\{ r^3 C^{TT,\,{\rm app}}(r) \}
=\frac{1}{2\pi^2} \int_0^\infty \! dk \, P(k)
 \left[ f^2(k)k^2 r \cos kr+\{ 2f^2(k)+g^2(k) \} k\sin kr
 \right].
\label{CRTTAPP}
\end{eqnarray}
Integrating by parts and applying the Fourier sine formula, 
we obtain a first-order differential equation for $P(k)$, 
\begin{eqnarray}
&&-k^2f^2(k)P'(k)+\left[ -2k^2f(k)f'(k)+kg^2(k) \right] P(k)
\nonumber \\
&&\hspace{2cm}
=4\pi \int_0^\infty \! dr \,
 \frac{1}{r} \frac{\partial}{\partial r} 
 \{ r^3 C^{TT,\,{\rm app}}(r) \} \sin kr
 \equiv S^{TT}(k).
\label{FORMULATT}
\end{eqnarray}
This is the basic equation for the inversion 
of the TT angular power spectrum 
to the primordial curvature perturbation spectrum. 
Note that $f(k)$ is an oscillatory function 
and the above differential equation is singular at $f(k)=0$. 
However, this can be regarded as an advantage. 
Namely, we can find the values of $P(k)$ at zero points of $f(k)$, 
say $k=k_s$, as
\begin{eqnarray}
P(k_s)=\frac{S^{TT}(k_s)}{k_s \, g^2(k_s)} \quad {\rm for} \quad f(k_s)=0,
\label{BCTT}
\end{eqnarray}
assuming that $P'(k)$ is finite at $k=k_s$. 
Thus, without solving the differential equation, 
it is possible to know a rough overall feature of the spectrum. 
We can then solve Eq.~(\ref{FORMULATT}) 
as a boundary value problem between the singularities. 

Now we consider the case of the polarization. 
Like Eq.~(\ref{CRTT}), we introduce the following quantities: 
\begin{eqnarray}
\tilde{C}^{EE}(\theta)
&\equiv& \sum_{\ell=0}^{\infty} \frac{2\ell+1}{4\pi}
         \frac{(\ell-2)!}{(\ell+2)!} \, C^{EE}_\ell P_\ell(\cos\theta),
\label{CREE} \\
\tilde{C}^{TE}(\theta)
&\equiv& \sum_{\ell=0}^{\infty} \frac{2\ell+1}{4\pi}
         \sqrt{\frac{(\ell-2)!}{(\ell+2)!}} \, C^{TE}_\ell P_\ell(\cos\theta).
\label{CRTE}
\end{eqnarray}
These are defined in such a way that the factors 
$(\ell+2)!/(\ell-2)!$ and $\sqrt{(\ell+2)!/(\ell-2)!}$ 
in $C^{EE}_\ell$ and $C^{TE}_\ell$, respectively, 
are canceled out so that we can use the formula (\ref{BESSEL}). 
Note that they are not the conventional angular correlation functions 
but just quantities that are convenient for inversion. 
Substituting Eqs.~(\ref{CLEEAPP}) and (\ref{CLTEAPP}) 
into Eqs.~(\ref{CREE}) and (\ref{CRTE}), respectively, 
in the small-scale limit $r \ll d$, we have 
\begin{eqnarray}
r\tilde{C}^{EE,\,{\rm app}}(r)
&=& \frac{1}{2\pi^2} \int_0^\infty \! dk \,
    kh^2(k)P(k) \sin kr,
\label{CREEAPP} \\
r\tilde{C}^{TE,\,{\rm app}}(r)
&=& \frac{1}{2\pi^2} \int_0^\infty \! dk \,
    kf(k)h(k)P(k) \sin kr\,.
\label{CRTEAPP}
\end{eqnarray}
We obtain expressions much simpler than that for the TT case. 
Applying the Fourier sine formula, 
we obtain the algebraic equations for $P(k)$, 
\begin{eqnarray}
kh^2(k)P(k)
&=& 4\pi \int_0^\infty \! dr \, r\tilde{C}^{EE,\,{\rm app}}(r) \sin kr
    \equiv S^{EE}(k),
\label{FORMULAEE} \\
kf(k)h(k)P(k)
&=& 4\pi \int_0^\infty \! dr \, r\tilde{C}^{TE,\,{\rm app}}(r) \sin kr
    \equiv S^{TE}(k).
\label{FORMULATE}
\end{eqnarray}
In this case, we can find $P(k)$ 
at the values of $k$ where $h(k) \ne 0$ for EE, and at those 
where $f(k) \ne 0$ and $h(k) \ne 0$ for TE 
without solving a differential equation. 
The equations (\ref{FORMULAEE}) and (\ref{FORMULATE}) 
are the basic inversion formulas for the EE and TE angular power spectra, 
respectively. 

In practice, however, we find that numerical errors become large 
around the singularities of Eq.~(\ref{FORMULAEE}) or (\ref{FORMULATE}), 
where the prefactor of $P(k)$ vanishes,
if we use either of them for inversion. 
We encountered a similar numerical problem in the TT case 
when we used Eq.~(\ref{FORMULATT}) for inversion, 
particularly when observational errors in the CMB power spectrum 
were taken into account~\cite{KMSY04}. 
Here, as a remedy for this numerical problem, 
we propose a new method that combines the TT and EE formulas. 
Since the zero points of $f(k)$ and $h(k)$ are 
different from each other as shown in Fig.~\ref{TF}, 
we expect this to suppress numerical errors 
around the zero points of both $f(k)$ and $h(k)$. 

Multiplying Eq.~(\ref{FORMULAEE}) by some factor $\alpha$ 
(which we take to be independent of $k$ for simplicity), 
and adding it to Eq.~(\ref{FORMULATT}), 
we obtain the combined inversion formula as 
\begin{eqnarray}
-k^2f^2(k)P'(k)+\left[ -2k^2f(k)f'(k)+kg^2(k)+\alpha kh^2(k) \right] P(k)
=S^{TT}(k)+\alpha S^{EE}(k),
\label{FORMULACOM}
\end{eqnarray}
and the boundary conditions are similarly given 
by the values of $P(k)$ at zero points of $f(k)$ as 
\begin{eqnarray}
P(k_s)=\frac{S^{TT}(k_s)+\alpha S^{EE}(k_s)}
            {k_s \left[ g^2(k_s)+\alpha h^2(k_s) \right]}
\quad {\rm for} \quad f(k_s)=0,
\label{BCCOM}
\end{eqnarray}
assuming that $P'(k)$ is finite at $k=k_s$. 
The contribution of EE is controlled by the parameter $\alpha$. 
If we take an appropriate value of $\alpha$ 
so that the contribution of EE is comparable to that of TT, 
the solution of Eq.~(\ref{FORMULACOM}) becomes numerically stable 
even around the singularities 
because the contribution of EE dominates 
near the singularities of TT given by $f(k)=0$, and vice versa. 
We investigate this in the next section. 

We do not use the TE formula, Eq.~(\ref{FORMULATE}), 
because it is singular not only at $h(k)=0$ but also at $f(k)=0$ 
where the TT formula is also singular~\footnote{
By combining Eq.~(\ref{FORMULATT}) 
and the derivative of Eq.~(\ref{FORMULATE}) with respect to $k$, 
we could obtain a non-singular formula for $P(k)$ 
in which the prefactor of $P(k)$ is non-vanishing. 
It turned out, however, 
that their performance was inferior to the present approach.}. 
Perhaps the TE spectrum can be used as a consistency check 
when we have accurate temperature and polarization maps at hand. 
However, here we do not discuss it.

\begin{figure}
\begin{center}
\includegraphics[width=10cm]{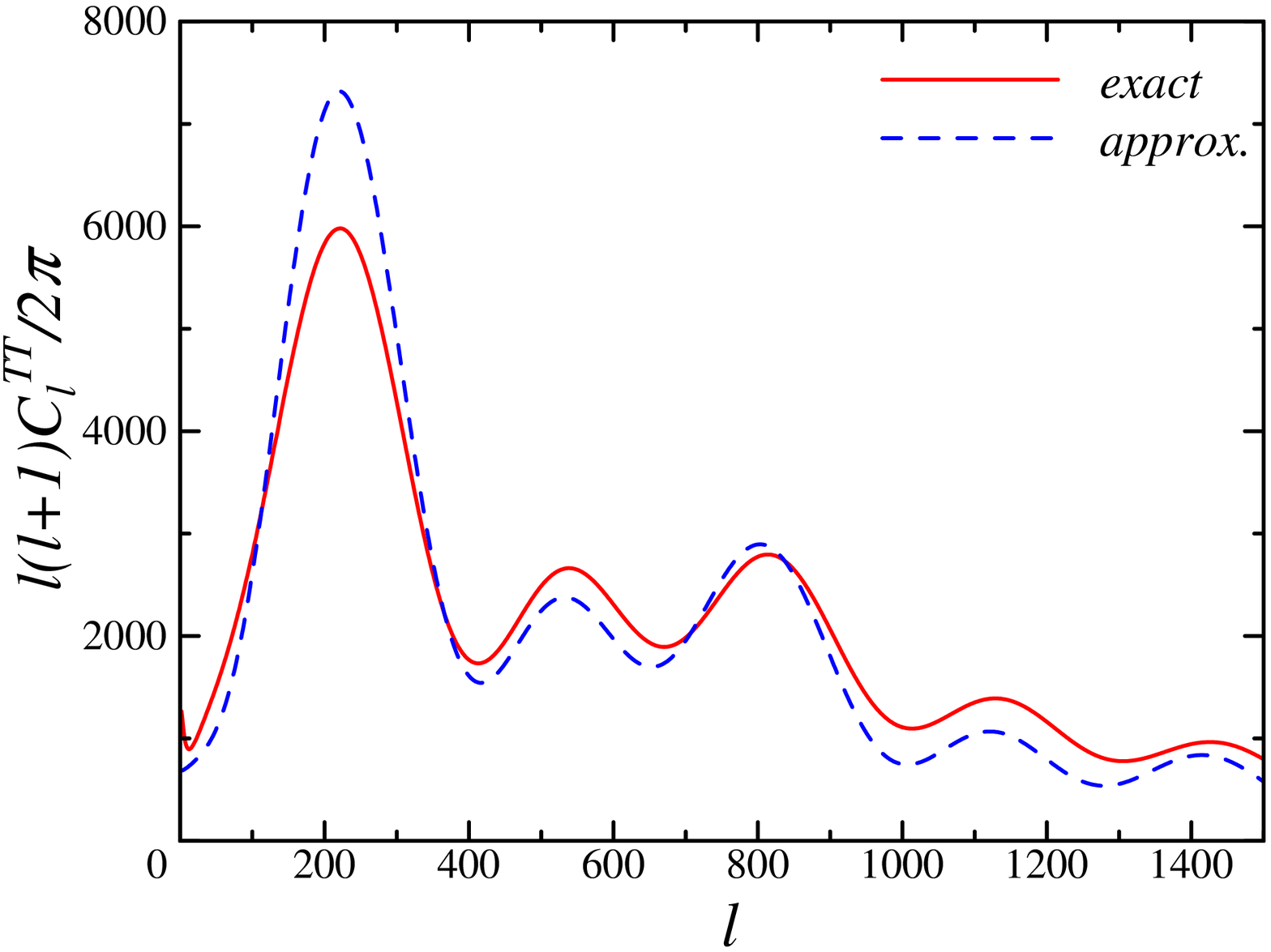}
\includegraphics[width=10cm]{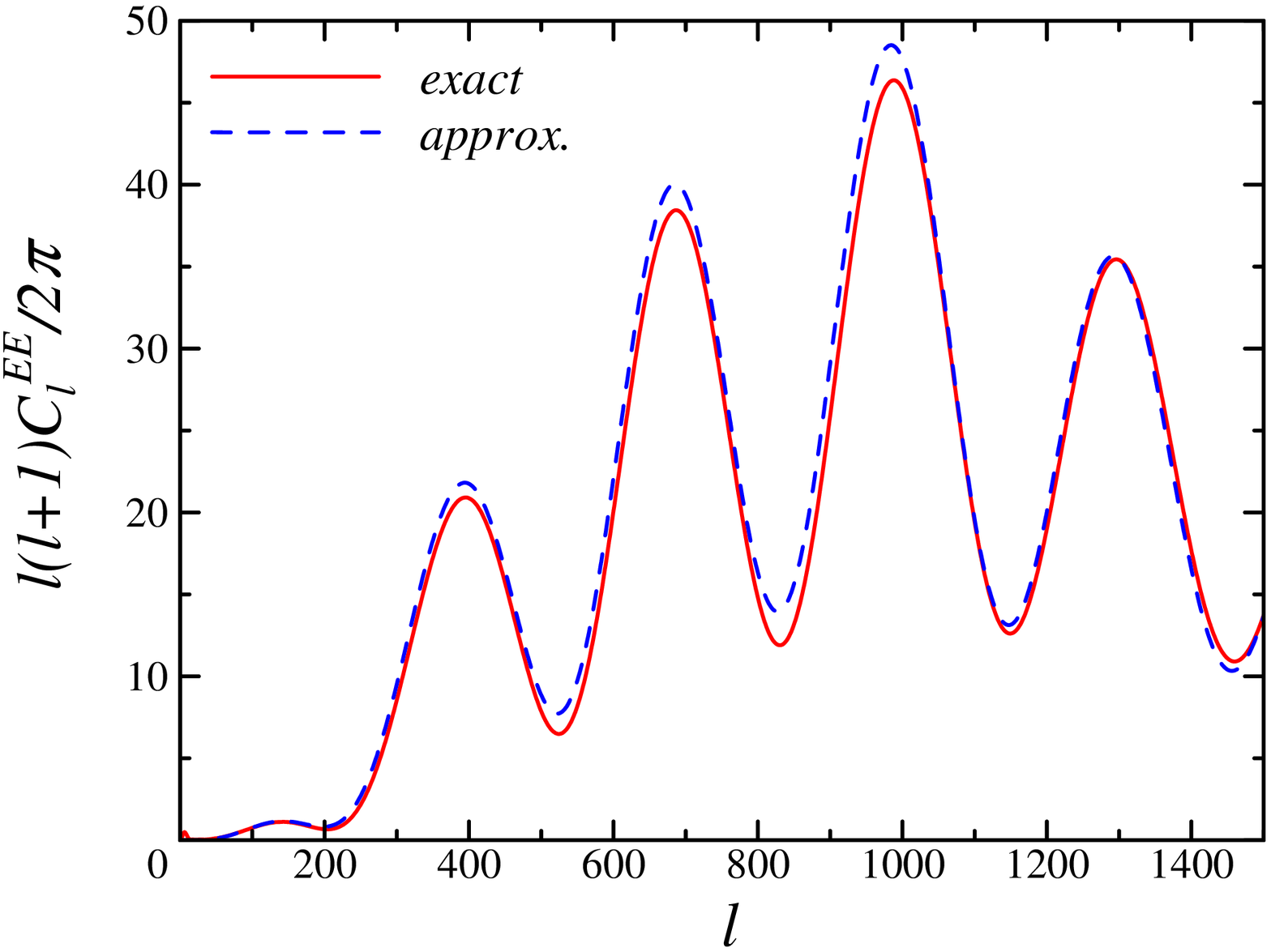}
\caption{Comparison of 
the exact spectrum $C^{XX,\,{\rm ex}}_\ell$ ({\it solid curves}) 
with the approximated one $C^{XX,\,{\rm app}}_\ell$ ({\it dashed curves}) 
for TT ({\it top panel}) and EE ({\it bottom panel}). 
The cosmological parameters are the same as those in Fig.~\ref{TF}, 
and $P(k)$ is assumed to be scale invariant. 
The relative errors of $C^{XX,\,{\rm app}}_\ell$ 
are as large as $20-30\%$ for both TT and EE. 
\label{CLAPP}}
\end{center}
\end{figure}

\begin{figure}
\begin{center}
\includegraphics[width=10cm]{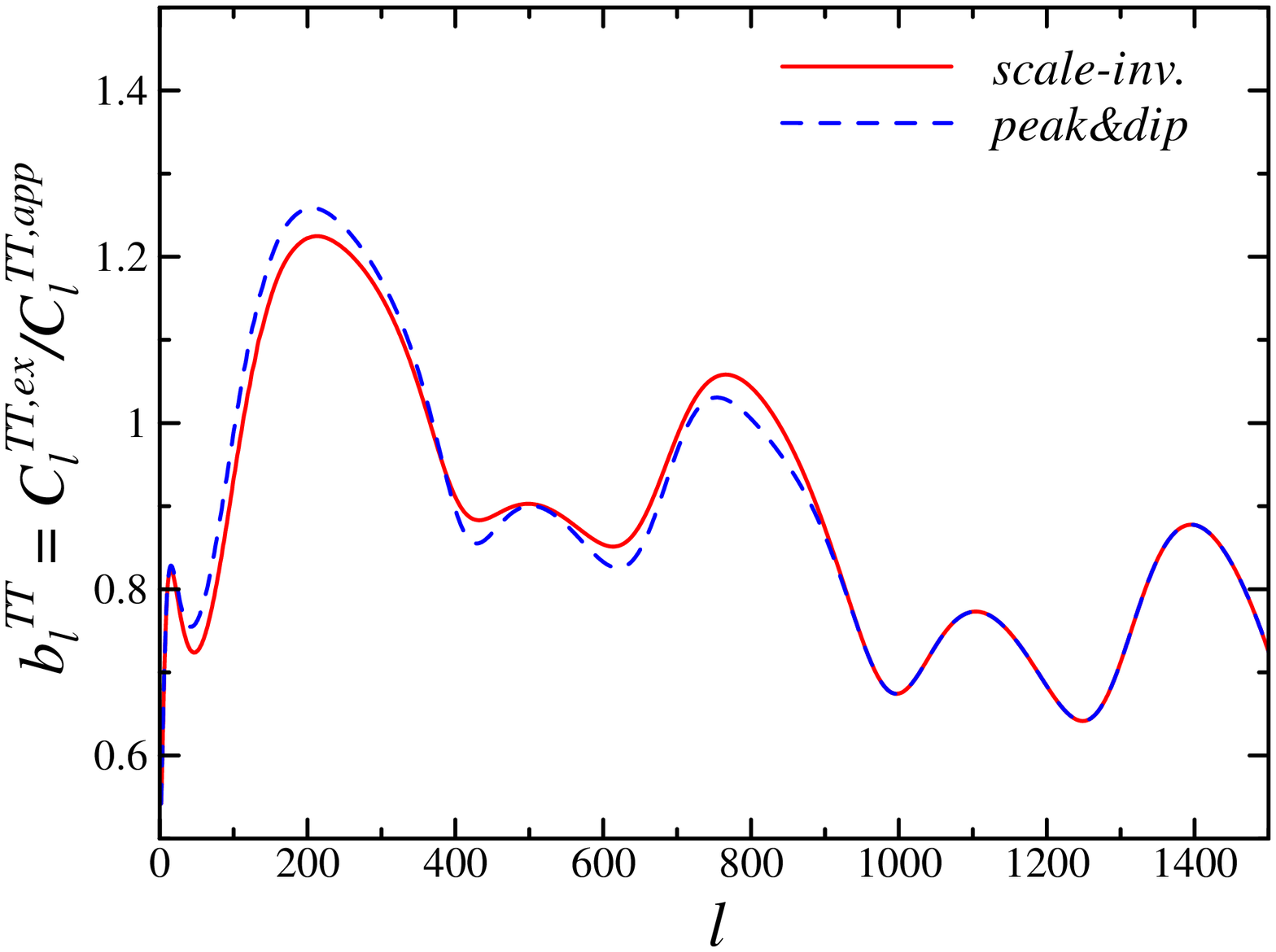}
\includegraphics[width=10cm]{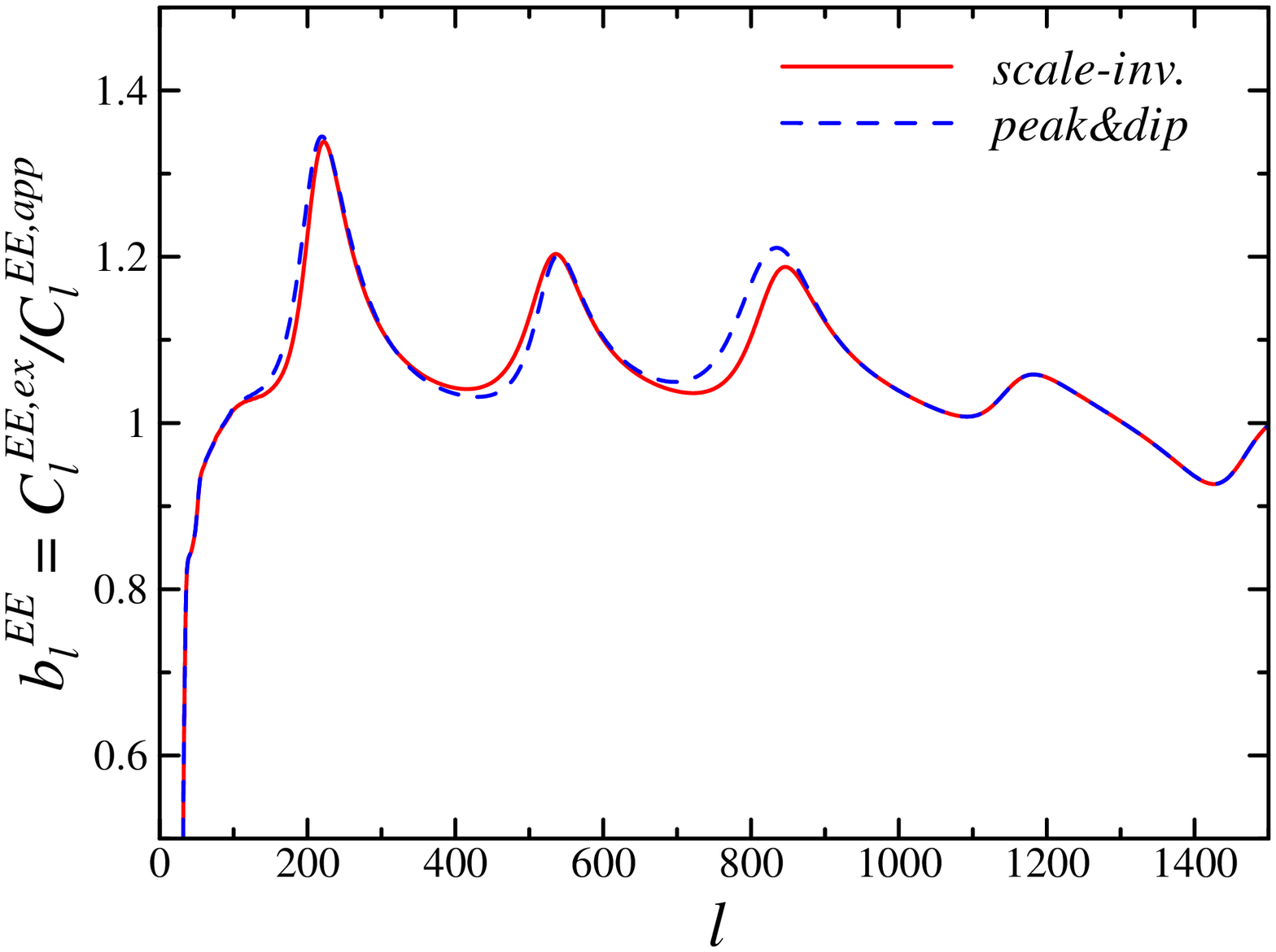}
\caption{Comparison of $b^{XX}_\ell$ 
between the case of a scale-invariant $P(k)$ ({\it solid curves}), 
and that of $P(k)$ with a peak and a dip ({\it dashed curves}), 
for TT ({\it top panel}) and EE ({\it bottom panel}). 
The cosmological parameters are the same as those in Fig.~\ref{TF}. 
The relative differences of $b^{XX}_\ell$ between these two cases 
are about $5\%$ for both TT and EE, 
implying the weak dependence of $b^{XX}_\ell$ on $P(k)$. 
\label{FL}}
\end{center} 
\end{figure}

For either TT or EE, 
the approximated spectrum $C^{XX,\,{\rm app}}_\ell$ 
($X=T$ or $E$) used to obtain the inversion formula 
has relative errors as large as about $20-30\%$ 
when compared with the exact one $C^{XX,\,{\rm ex}}_\ell$. 
We show $C^{XX,\,{\rm ex}}_\ell$ and $C^{XX,\,{\rm app}}_\ell$ 
for TT and EE in Fig.~\ref{CLAPP}. 
Of course, since the observed spectrum $C^{XX,\,{\rm obs}}_\ell$ 
is to be compared with $C^{XX,\,{\rm ex}}_\ell$, 
it is necessary to correct the errors. 
For this purpose, we first define the ratio, 
\begin{eqnarray}
b^{XX}_\ell \equiv
\frac{C^{XX,\,{\rm ex}}_\ell}{C^{XX,\,{\rm app}}_\ell}\,.
\label{RATIOXX}
\end{eqnarray}
Interestingly, this ratio is found to be almost independent of $P(k)$ 
not only for TT~\cite{MSY03} but also EE. 
Figure~\ref{FL} shows $b^{TT}_\ell$ and $b^{EE}_\ell$ 
for a scale-invariant $P(k)$ and a spectrum with a peak and a dip. 
In this case, the difference is found to be less than about $5\%$. 
We use this property of $b^{XX}_\ell$ to perform an iteration 
of inversion as previously done in the TT case~\cite{MSY02,MSY03}. 

For self-containedness, let us recapitulate the iterative procedure. 
First we calculate 
$b^{XX,\,(0)}_\ell=C^{XX,\,{\rm ex}(0)}_\ell/C^{XX,\,{\rm app}(0)}_\ell$, 
for a fiducial primordial spectrum $P^{(0)}(k)$ 
such as a scale-invariant one. 
Then we divide the observed spectrum 
$C^{XX,\,{\rm obs}}_\ell$ by $b^{XX,\,(0)}_\ell$; 
$C^{XX,\,{\rm obs}}_\ell/b^{XX,\,(0)}_\ell$. 
This gives a better guess to $C^{XX,\,{\rm app}}_\ell$. 
We use it to evaluate 
$S^{XX}(k)$ in the right-hand side of Eq.~(\ref{FORMULACOM}) 
and solve this equation. 
Let us denote the reconstructed primordial spectrum 
from $C^{XX,\,{\rm obs}}_\ell/b^{XX,\,(0)}_\ell$ by $P^{(1)}(k)$. 
Then we can repeat the same procedure and obtain $P^{(2)}(k)$ 
from $C^{XX,\,{\rm obs}}_\ell/b^{XX,\,(1)}_\ell$, 
where $b^{XX,\,(1)}_\ell$ is calculated from $P^{(1)}(k)$. 
Repeating this procedure as many times as it is necessary, 
we can find $P(k)$ with much better accuracy. 
The $n$-th iteration is summarized as 
\begin{eqnarray}
\cdots \Rightarrow \,\, P^{(n-1)}(k)
\,\, \rightarrow \,\, b_\ell^{XX,\,(n-1)}
                     =\frac{C^{XX,\,{\rm ex}(n-1)}_\ell}
                           {C^{XX,\,{\rm app}(n-1)}_\ell}
\,\, \rightarrow \,\, \frac{C^{XX,\,{\rm obs}}_\ell}{b_\ell^{XX,\,(n-1)}}
\,\, \Rightarrow \,\, P^{(n)}(k)
\,\, \rightarrow \cdots.
\label{ITETT}
\end{eqnarray}
We confirm that this iteration converges within a few times 
as discussed in the next section. 

\section{Numerical Tests} \label{TESTS}

To test our inversion method including the CMB polarization, 
we perform the reconstruction of $P(k)$ from $C^{XX}_\ell$ 
for several $P(k)$ whose shapes are given by hand. 
As mentioned before, we assume here an ideal situation in which 
the CMB data contain no observational error. 
We use $C^{XX}_\ell$ up to $\ell_{\rm max}=1500$ for both TT and EE. 
As a fiducial spectrum, $P^{(0)}(k)$, for the iteration, 
we choose a scale-invariant spectrum $k^3P(k)=\mbox{const.}$, 
and assume the cosmological parameters as 
$h=0.70$, $\Omega_b=0.050$, $\Omega_\Lambda=0.70$, 
$\Omega_m=0.30$, and $\tau=0.20$. 
In this case, the positions of the singularities for $f(k)=0$ are at 
$kd \simeq 70$, $430$, $680$, $1030$, $\cdots$, 
and for $h(k)=0$ are at $kd \simeq 230$, $560$, $860$, $1180$, $\cdots$, 
where $d \simeq 1.36\times10^4\,{\rm Mpc}$. 
We solve Eq.~(\ref{FORMULACOM}) as a boundary value problem 
between the first and fourth TT singularities for $f(k)=0$. 
Hence the range of the reconstructed $P(k)$ is $70 \le kd \le 1030$, 
where $kd$ roughly corresponds to $\ell$. 
When proceeding from the $(n-1)$-th to $n$-th iteration, 
we need $P^{(n-1)}(k)$ up to $k_{\rm max}d \simeq 2\ell_{\rm max}$ 
to calculate $b^{XX,\,(n-1)}_\ell$. 
Hence we assume the scale-invariance outside of the reconstructed range. 

\begin{figure}
\begin{center}
\includegraphics[width=10cm]{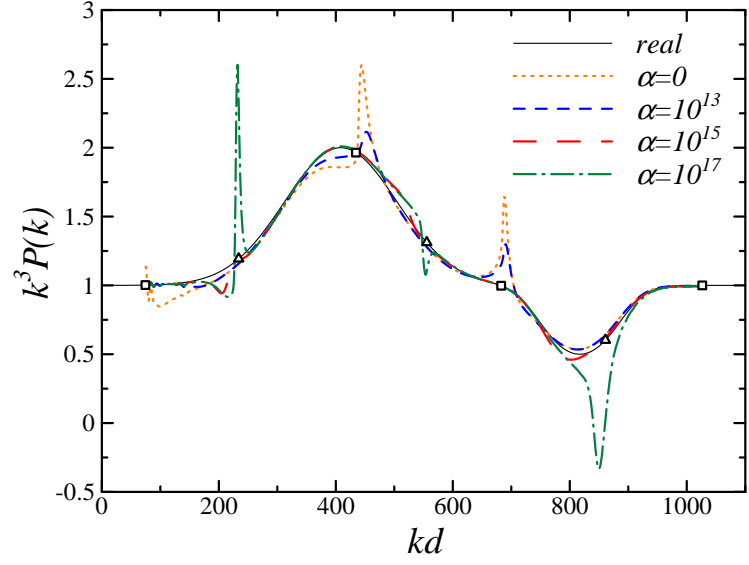}
\caption{Dependence of the reconstructed $P(k)$ on the parameter $\alpha$ 
which controls the contribution of EE relative to TT, 
for the presumed form of $P(k)$ given 
by Eq.~(\ref{PKPD}) in the text. 
The cosmological parameters are assumed to be known. 
The result of reconstruction is shown 
for $\alpha=0$, $10^{13}$, $10^{15}$, and $10^{17}$. 
The horizontal axis $kd$ roughly corresponds to $\ell$. 
The singularities $f(k)=0$ are denoted by $\square$, 
and $h(k)=0$ by $\triangle$. 
In the case of $\alpha=10^{15}$, 
for which the contributions of TT and EE are comparable, 
the errors are substantially suppressed, especially near the singularities. 
\label{ALPHA}}
\end{center}
\end{figure}

First we examine the dependence of the reconstructed $P(k)$ 
on the parameter $\alpha$ appeared in Eq.~(\ref{FORMULACOM}), 
which controls the contribution of EE, 
assuming that the cosmological parameters are known. 
We assume that $P(k)$ has a peak and a dip expressed as 
\begin{eqnarray}
k^3 P(k)
=A \left\{1+a_1\exp\left[ -\frac{(k-k_1)^2}{\sigma_1{}^2} \right]\right\}
   \left\{1+a_2\exp\left[ -\frac{(k-k_2)^2}{\sigma_2{}^2} \right]\right\}^{-1},
\label{PKPD}
\end{eqnarray}
and we take $a_1=a_2=1.0$, 
$k_1=0.03\,{\rm Mpc}^{-1}$, $k_2=0.06\,{\rm Mpc}^{-1}$, 
$\sigma_1=0.01\,{\rm Mpc}^{-1}$, and $\sigma_2=0.005\,{\rm Mpc}^{-1}$. 
Figure~\ref{ALPHA} depicts the reconstructed $P(k)$ 
with four different values of $\alpha$. 
The case $\alpha=0$ corresponds to 
the result of the original cosmic inversion method. 
We show only $P^{(1)}(k)$ for each $\alpha$, 
which means that no iteration is performed, 
to see the effect of the contribution of EE explicitly. 
We find that the reconstructed $P(k)$ is in a good agreement 
with the original one even near the singularities for $\alpha=10^{15}$, 
while spurious sharp features appear near the TT singularities 
for $\alpha$ much smaller than $10^{15}$ 
and those appear near the EE singularities 
for $\alpha$ much larger than $10^{15}$. 
This behavior is reasonable 
because the contribution of EE is found to be of the same order of TT 
for $\alpha \sim 10^{14-15}$ 
by comparing their magnitudes in the source terms of Eq.~(\ref{FORMULACOM}). 
Therefore, we adopt $\alpha=10^{15}$ in the following reconstructions. 
It may be noted that 
the origin of this number, $\sim10^{14-15}$, 
comes dominantly from the fact that the transfer function $h(k)$ for EE 
is intrinsically smaller than the transfer functions $f(k)$ and $g(k)$ 
for TT by a factor $\sim 10^{-7}$ (see Fig.~\ref{TF}) 
and their squares are contained 
in the left-hand side of Eq.~(\ref{FORMULACOM}). 

We perform the inversion for several shapes of $P(k)$, 
assuming that the cosmological parameters are known. 

\begin{figure}
\begin{center}
\includegraphics[width=10cm]{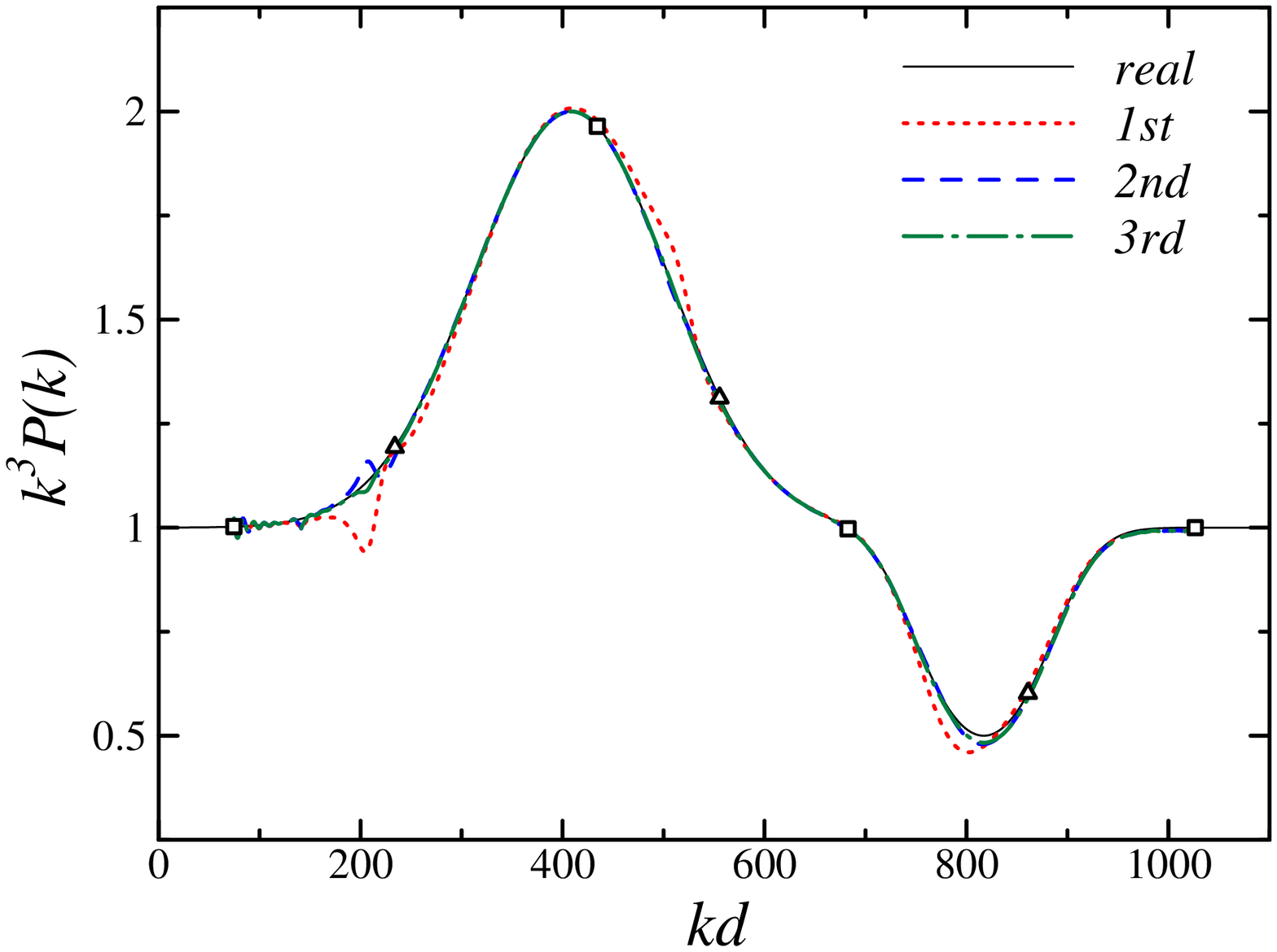}
\includegraphics[width=10cm]{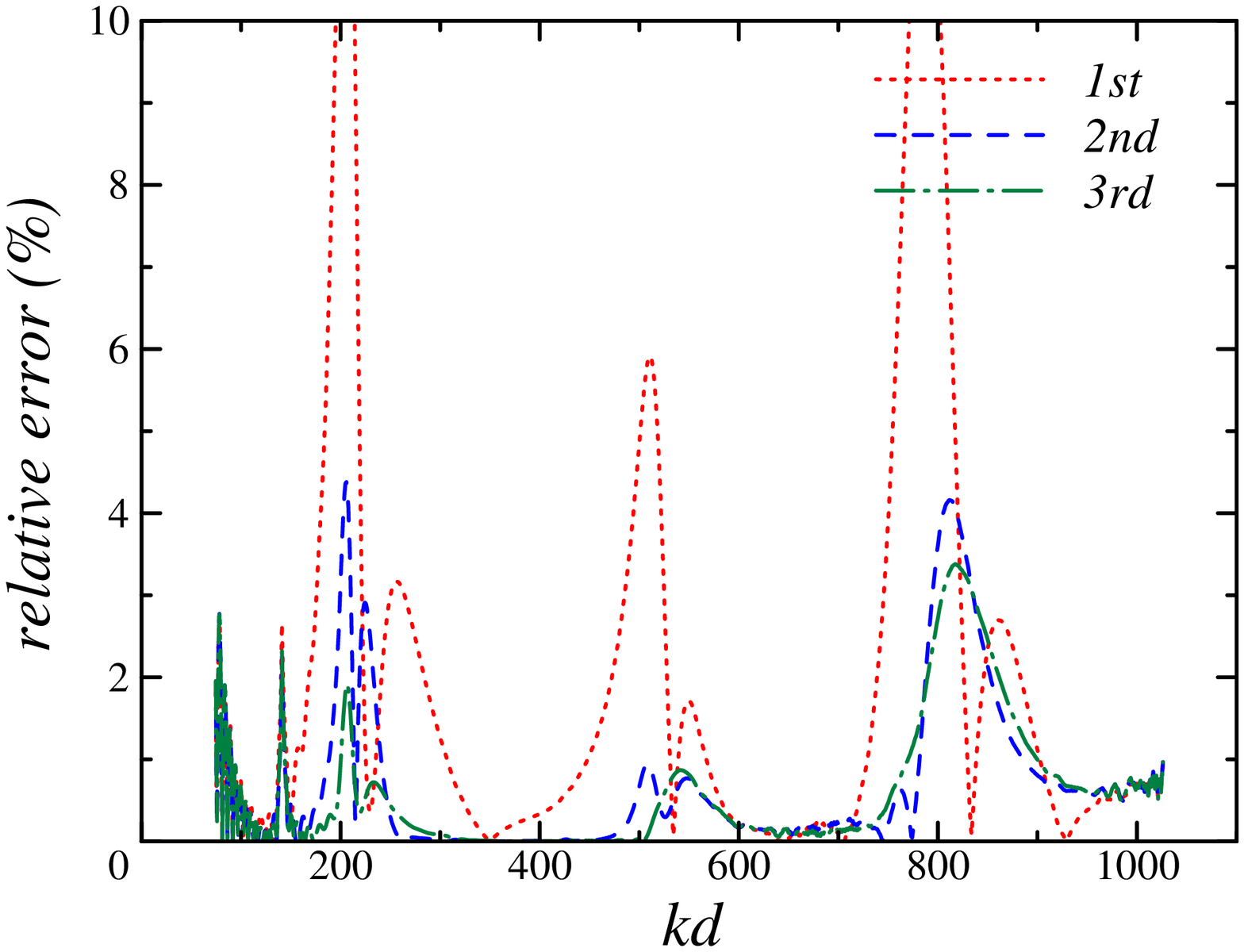}
\caption{Reconstructed spectrum for $P(k)$ with a peak and a dip, 
given by Eq.~(\ref{PKPD}) in the text. 
The top panel shows the results after each round of iteration, 
and the bottom panel shows their relative errors, 
$|P^{(n)}(k)-P^{\rm real}(k)|/P^{\rm real}(k)$. 
The singularities $f(k)=0$ are denoted by $\square$, 
and $h(k)=0$ by $\triangle$. 
\label{REPD}}
\end{center}
\end{figure}

\begin{figure}
\begin{center}
\includegraphics[width=10cm]{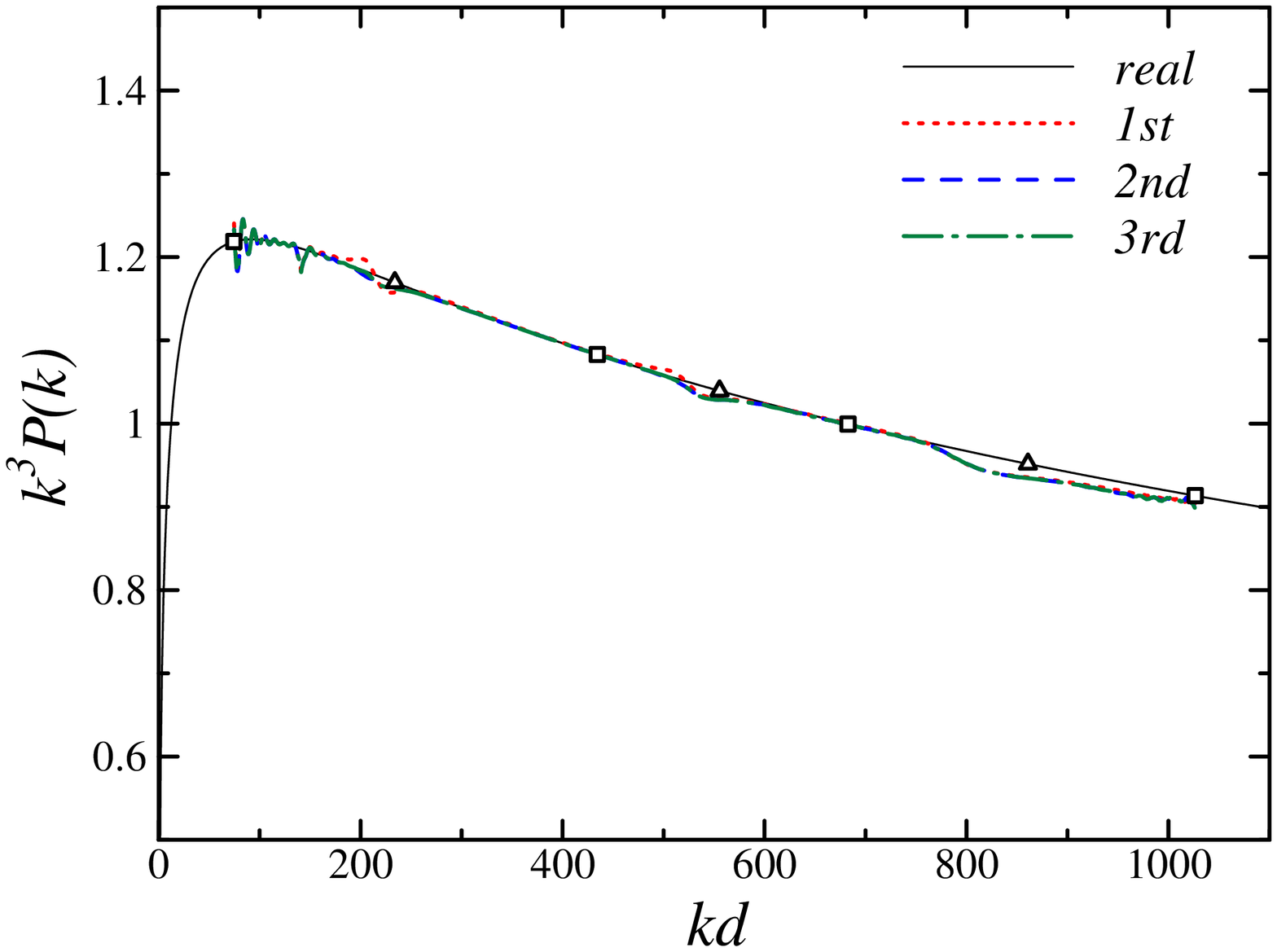}
\includegraphics[width=10cm]{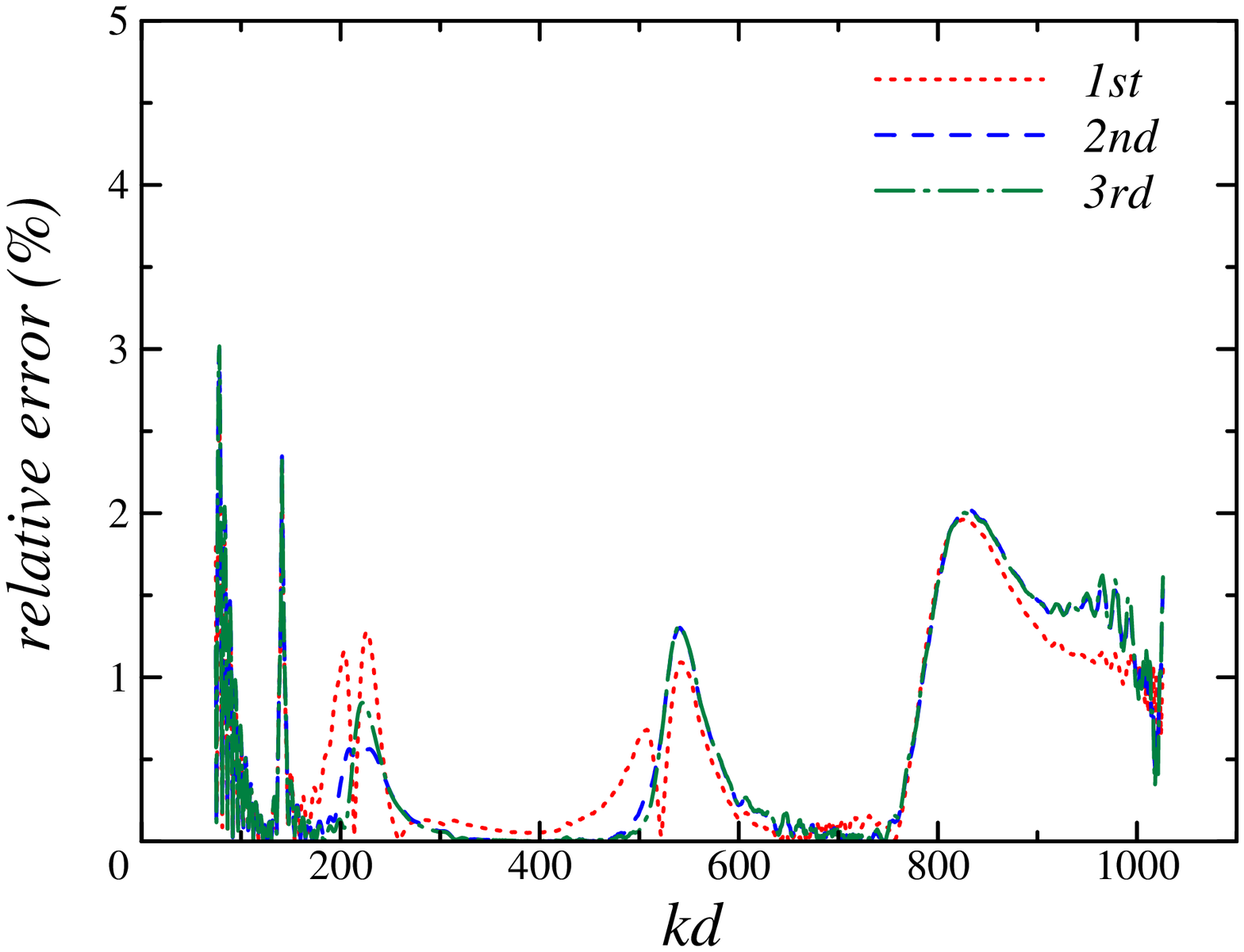}
\caption{The same as Fig.~\ref{REPD}, 
but for $P(k)$ with a running spectral index, 
given by Eq.~(\ref{PKRUN}) in the text. 
In this case, the errors are already less than $3\%$ 
at the first round of iteration, 
and they remain stably at that level 
after the second and third rounds of iteration. 
\label{RERUN}}
\end{center}
\end{figure}

\begin{figure}
\begin{center}
\includegraphics[width=10cm]{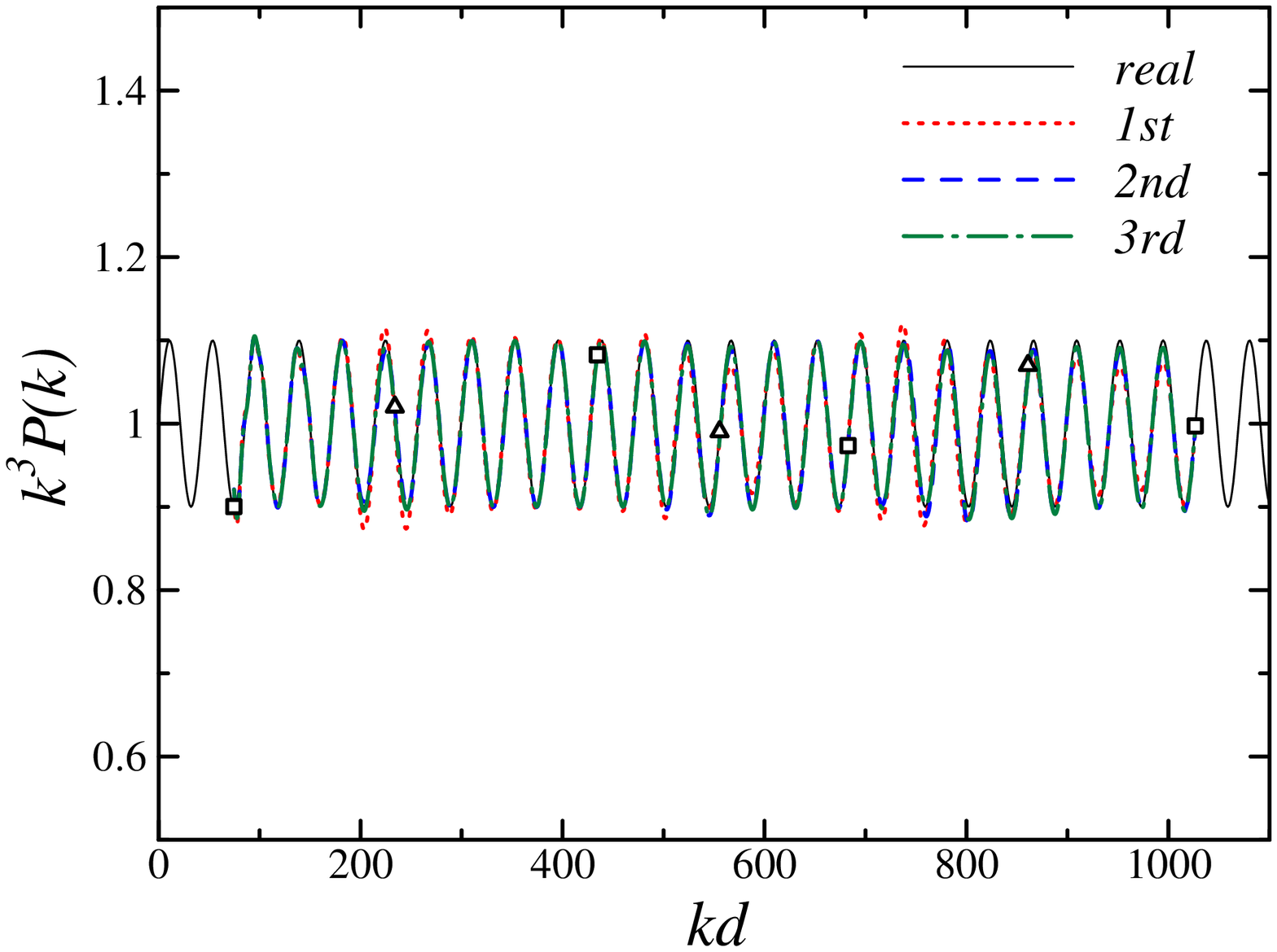}
\includegraphics[width=10cm]{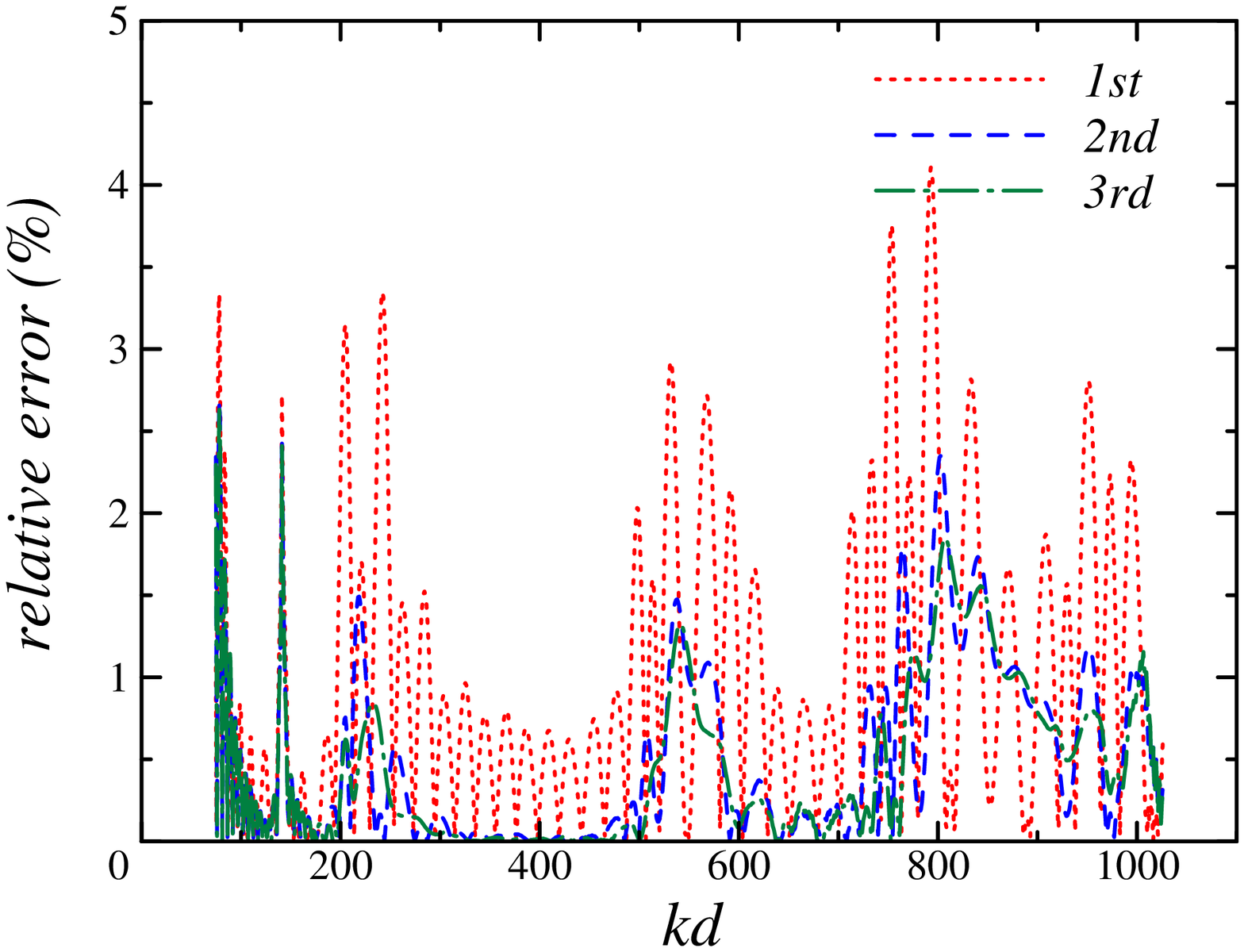}
\caption{The same as Fig.~\ref{REPD}, 
but for $P(k)$ with an oscillatory component, 
given by Eq.~(\ref{PKOSC}) in the text. 
The results are quite impressive 
because such a small oscillation like this 
can be reconstructed with the errors less than $3\%$. 
\label{REOSC}}
\end{center}
\end{figure}

\begin{list}{}{}
\item[(i)] {\it Spectrum with a peak and a dip}

Figure~\ref{REPD} shows the results of the iteration 
for $P(k)$ given by Eq.~(\ref{PKPD}). 
We see that the iteration converges within a few times 
and the relative errors are reduced to about a few percent level. 
In~\cite{MSY03}, it has been shown that in the case of 
using the TT spectrum only, which is identical to $\alpha=0$ here, 
one can reconstruct $P(k)$ with about a few percent errors by iteration, 
cutting off spurious sharp features appeared near the singularities 
and interpolating the spectrum smoothly at each step of iteration. 
But with our new method, it became unnecessary 
to perform this artificial modification of the spectrum during iteration. 

\item[(ii)] {\it Spectrum with a running spectral index}

Figure~\ref{RERUN} shows the results 
for $P(k)$ with a running spectral index given by~\cite{KT95} 
\begin{eqnarray}
k^3 P(k)
=A \left( \frac{k}{k_0} \right)^{n(k)-1}, \quad
n(k)
=n(k_0)+\frac{dn}{d\ln k}\ln\left( \frac{k}{k_0} \right).
\label{PKRUN}
\end{eqnarray}
We take $k_0=0.05\,{\rm Mpc}^{-1}$, $n(k_0)=0.8$, and $dn/d\ln k=-0.05$. 

\item[(iii)] {\it Spectrum with an oscillatory component}

Figure~\ref{REOSC} shows the results 
for $P(k)$ with an oscillatory component given by 
\begin{eqnarray}
k^3 P(k)
=A \left[ 1+a_0\sin\left( \frac{k}{k_0} \right) \right].
\label{PKOSC}
\end{eqnarray}
This form is motivated by possible signatures of 
trans-Planckian physics~\cite{MB01,BM01,MB03}. 
We take $a=0.1$ and $k_0=5\times10^{-4}\,{\rm Mpc}^{-1}$. 

\end{list}

The results are quite impressive because we can reconstruct 
not only a global shape such as a running spectral index, 
but also a fine feature such as a small oscillation 
as given by Eq.~(\ref{PKOSC}) with only a few percent errors. 

\section{Novel Method to Constrain Cosmological Parameters} \label{PARAMETERS}

We also examine how the shape of reconstructed $P(k)$ changes 
if we adopt the values of the cosmological parameters 
different from the assumed values, 
namely, those used in making a mock sample of $C^{XX}_\ell$. 
We vary each cosmological parameter 
in the range of about $2\sigma$ of its current best-fit value, 
that is, $0.60 \le h \le 0.80$, $0.040 \le \Omega_b \le 0.060$, 
and $0.60 \le \Omega_\Lambda \le 0.80$, respectively, 
with the others fixed in the case of a scale-invariant $P(k)$. 
We do not vary $\tau$ since it affects only the normalization 
except for the spectrum on large scales. 

\begin{figure}
\begin{center}
\includegraphics[width=8cm]{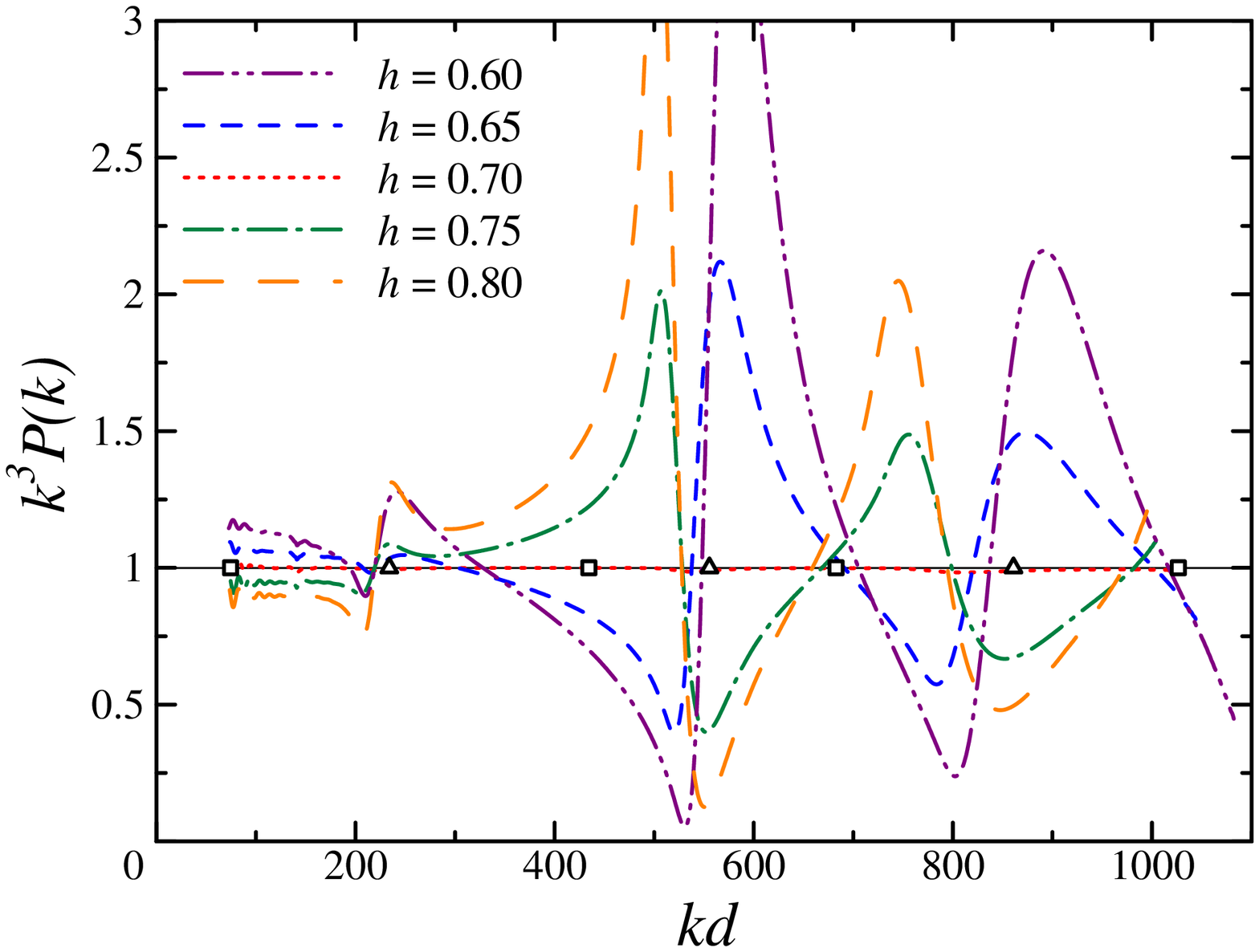}
\includegraphics[width=8cm]{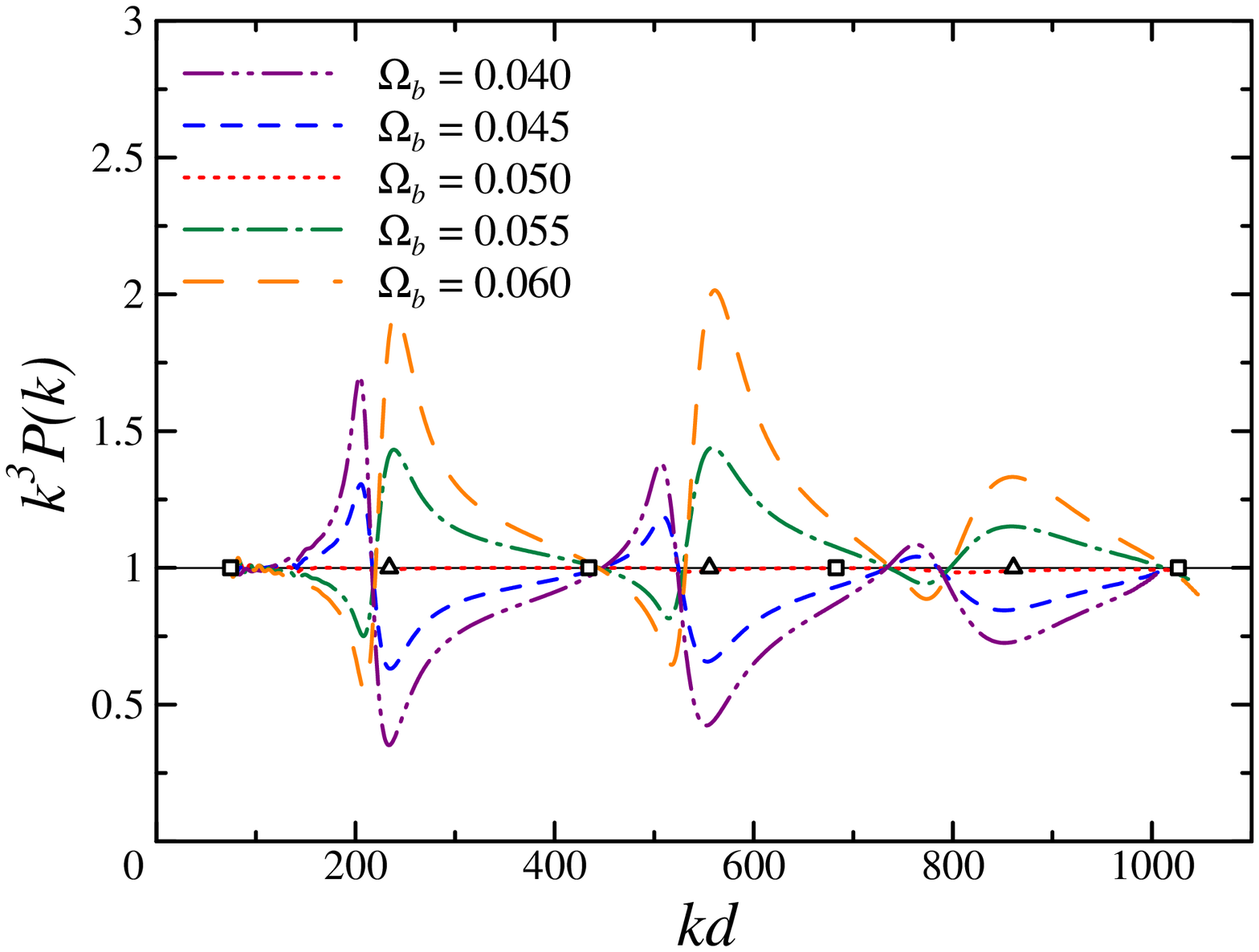}
\includegraphics[width=8cm]{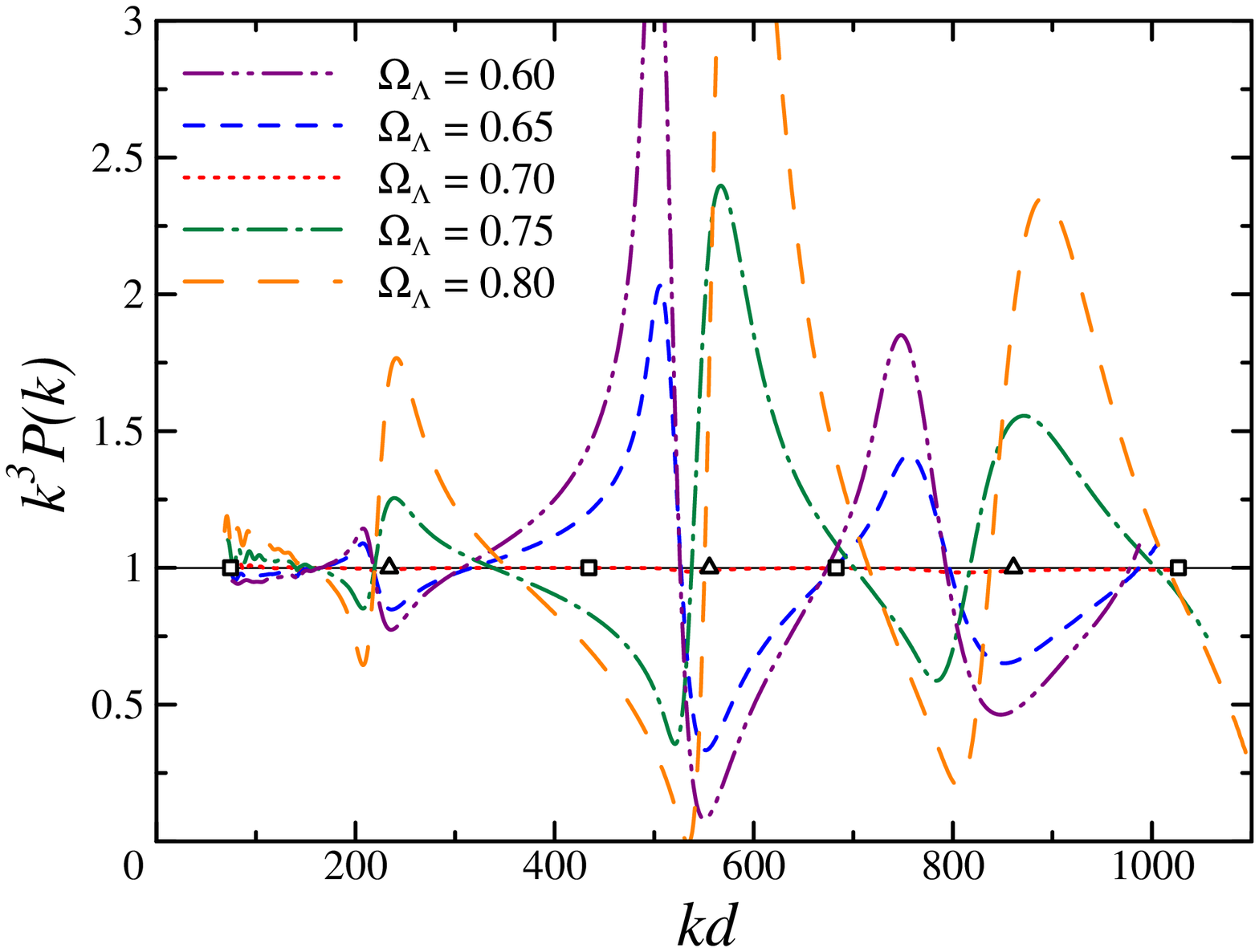}
\caption{Dependence of the reconstructed $P(k)$ 
on the cosmological parameters different from the presumed real values. 
We assume a scale-invariant $P(k)$ and the cosmological parameters as 
$h=0.70$, $\Omega_b=0.050$, $\Omega_\Lambda=0.70$, 
$\Omega_m=0.30$, and $\tau=0.20$. 
The top left panel shows the reconstruction by varying $h$ over 
$h=0.60$, $0.65$, $0.70$, $0.75$, and $0.80$. 
The top right panel shows the reconstruction by varying $\Omega_b$ 
over $\Omega_b=0.040$, $0.045$, $0.050$, $0.055$, and $0.060$. 
The bottom panel shows the reconstruction by varying $\Omega_\Lambda$ 
over $\Omega_\Lambda=0.60$, $0.65$, $0.70$, $0.75$, and $0.80$. 
In all the cases, the other parameters are fixed at the presumed values. 
Note that the value of $d$ and the positions of the singularities 
depend on the cosmological parameters. 
Plotted are the singularities 
($f(k)=0$ by $\square$ and $h(k)=0$ by $\triangle$) 
in the case of the presumed values of the cosmological parameters. 
\label{PARA}}
\end{center}
\end{figure}

\begin{figure}
\begin{center}
\includegraphics[width=10cm]{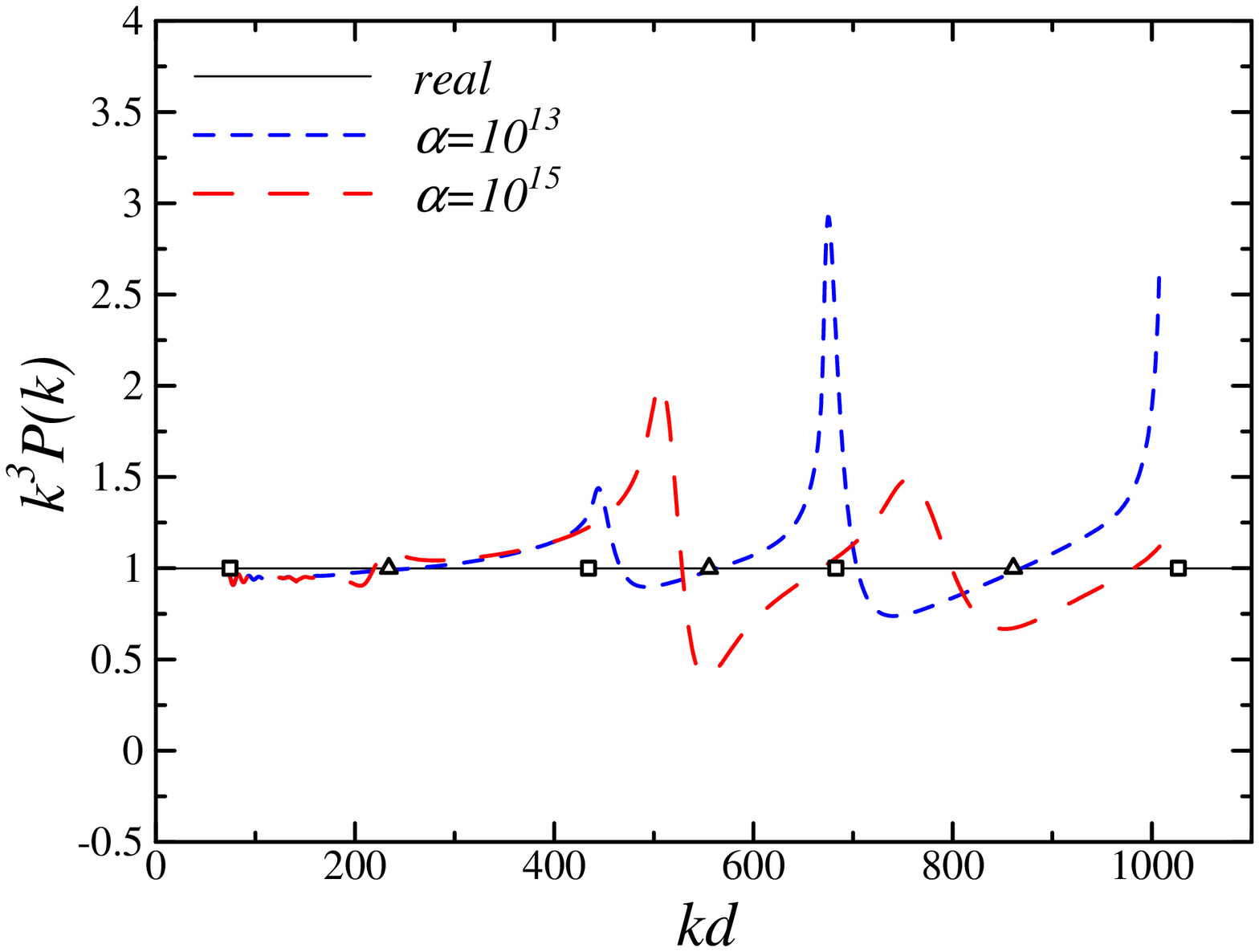}
\includegraphics[width=10cm]{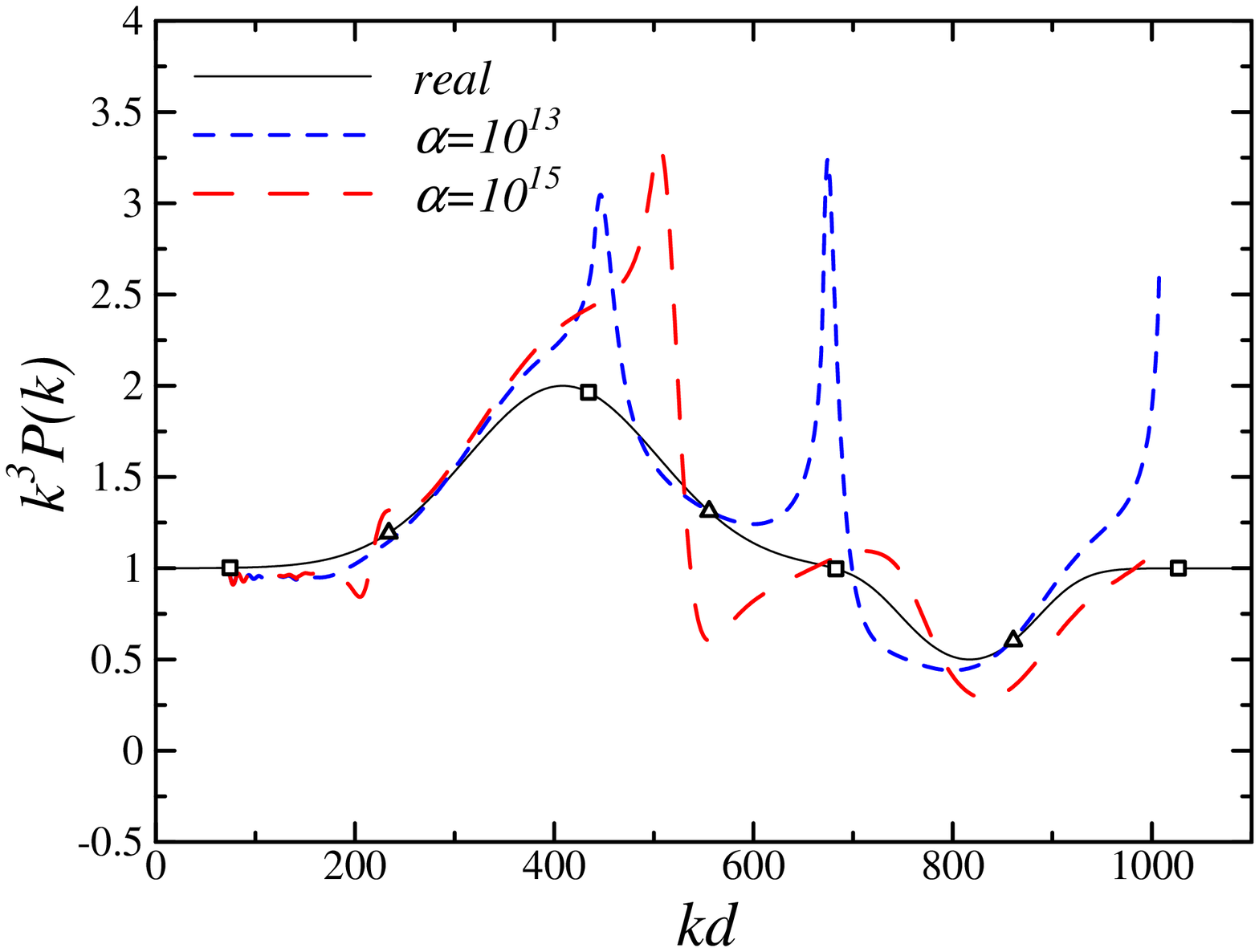}
\caption{Dependence of the reconstructed $P(k)$ on the parameter $\alpha$ 
in the case that one of the cosmological parameters is incorrect. 
Assumed values of the cosmological parameters 
are the same as those in Fig.~\ref{PARA}. 
Both are the cases that we use an incorrect value, $h=0.75$ 
in our reconstruction for $\alpha=10^{13}$ and $10^{15}$. 
The top panel shows the reconstructed spectra for a scale-invariant $P(k)$ 
and the bottom panel shows those for $P(k)$ with a peak and a dip 
given by Eq.~(\ref{PKPD}) in the text. 
Plotted are the singularities in the same way as Fig.~\ref{PARA}. 
\label{RESING}}
\end{center}
\end{figure}

The resultant $P^{(1)}(k)$ are shown in Fig.~\ref{PARA}. 
We see that if we use incorrect values of the cosmological parameters 
in our reconstruction procedure, 
the reconstructed $P(k)$ is deformed from its real shape 
near the EE singularities for $\alpha=10^{15}$. 
This is because the contribution of EE is slightly larger than that of TT. 
If we adopt a smaller value of $\alpha$ 
so that the contribution of TT is larger than that of EE, 
such a deformation appears near the TT singularities. 
In fact, this was the case studied in~\cite{MSY03} 
which corresponds to the extreme case $\alpha=0$ using only the TT spectrum. 
Figure~\ref{RESING} shows how the reconstructed $P(k)$ 
with incorrect values of the cosmological parameters 
changes depending on $\alpha$ for the case that $P(k)$ is scale invariant 
and that it has a peak and a dip given by Eq.~(\ref{PKPD}). 

In the previous case where we use only TT spectrum, 
even if we ended up with a spiky shape of the reconstructed $P(k)$ 
around some specific values of $k$, 
it is difficult to judge if it was a real feature or simply an artifact 
due to the incorrect choice of the cosmological parameters. 
In contrast, in the present case where we use both TT and EE spectra, 
we can invent a new method to constrain the cosmological parameters 
without assuming the shape of $P(k)$ from the following two observations. 
First, as shown in Fig.~\ref{ALPHA}, 
in the range of $10^{13} \lesssim \alpha \lesssim 10^{15}$, 
$P(k)$ is reconstructed rather well 
if we use the correct values of the cosmological parameters. 
On the other hand, if we use incorrect values in our reconstruction, 
we obtain a deformed shape of the reconstructed $P(k)$. 
Now, as shown in Fig.~\ref{RESING}, 
the location of the deformation in $k$-space 
depends on the value of $\alpha$ we use. 
That is, if we take a relatively large value of $\alpha$, 
say, $\alpha_1 \sim 10^{15}$, 
the deformation is most prominent around the EE singularities. 
On the other hand, if we use a smaller value, 
say, $\alpha_2 \sim 10^{13}$, 
the location of the deformation moves to the TT singularities. 
Thus unless we adopt the cosmological parameters 
which are close enough to the real values, 
the shape of $P_{\alpha_1}(k)$, 
namely the reconstructed spectrum for $\alpha=\alpha_1$, 
will be much different 
from the shape of $P_{\alpha_2}(k)$, that for $\alpha=\alpha_2$. 
Thus we may obtain constraints on the cosmological parameters 
by comparison of the shapes of $P_{\alpha_1}(k)$ and $P_{\alpha_2}(k)$. 
An intriguing point is that no assumption on the shape of $P(k)$ 
is necessary to derive the constraints. 

\section{Conclusion} \label{CONCLUSION}

We have derived the inversion formula 
including both the CMB temperature fluctuations and polarization, 
and examined the validity of our inversion method. 
Our method is based on a linear combination of inversion formulas 
for approximate temperature (TT) and polarization (EE) spectra 
in the form $\mbox{TT}+\alpha\mbox{EE}$, 
given by Eq.~(\ref{FORMULACOM}), with a constant parameter $\alpha$, 
combined with iteration that corrects errors in the approximate formulas. 

First we have examined the effect of the contribution of 
EE on the reconstructed $P(k)$. 
There are singularities in the TT and EE inversion formulas 
due to the zero points of transfer functions. 
We have found that large numerical errors in the reconstructed $P(k)$ 
around the singularities of the TT and EE inversion formulas, 
which appear when each formula is used independently, 
disappear for the choice of $\alpha \sim 10^{14-15}$, 
for which the contribution of EE becomes comparable to that of TT. 
Using such an appropriate value of $\alpha$, 
we have tested our method against several presumed spectral shapes of $P(k)$. 
As a result, 
we have confirmed that the relative errors 
between the reconstructed $P(k)$ and the original one 
are reduced to a few percent level after a few steps of iteration. 

We have also examined the case 
when the cosmological parameters are varied from the presumed real values. 
For the choice of $\alpha=10^{15}$, 
we have shown that the reconstructed $P(k)$ 
is substantially deformed from its real shape near the EE singularities, 
while it is substantially deformed near the TT singularities 
for much smaller values of $\alpha$. 
Since the errors near the singularities in the reconstructed $P(k)$ 
are relatively small for any $\alpha$ in the range of 
$10^{13} \lesssim \alpha \lesssim 10^{15}$, 
if the correct cosmological parameters are chosen, 
this result suggests a new method 
for constraining the cosmological parameters. 
Namely, by comparison of reconstructed $P(k)$ 
for different values of $\alpha$, 
we may constrain the cosmological parameters 
without any assumption on the shape of $P(k)$. 

In this paper, we have not taken account of possible observational errors. 
In reality, of course, observational errors are by no means negligible. 
As shown in~\cite{KMSY04}, in the case of using the TT spectrum of the WMAP, 
numerical errors are enhanced near the singularities, 
which made us impossible to obtain any information of $P(k)$ 
near the singularities. 
However, our new method that 
substantially suppresses the numerical errors around the singularities 
may not suffer much from this difficulty. 
In any case, it is important to investigate the effect 
of observational errors on our new method. 
It is also important to investigate the possibility 
of using our new inversion method to constrain the cosmological parameters 
by taking account of observational errors. 
Since we will inevitably obtain different shapes of $P(k)$ 
for different choices of $\alpha$ under the presence of observational errors, 
even if we knew and used the true cosmological parameters, 
it is necessary to develop a statistical criterion 
that tests the equivalence of the shapes of reconstructed $P(k)$. 
It is also necessary to extend our method to include 
the contribution of tensor-type perturbations, 
which is probed by the B-mode polarization. 
Their contribution may be appreciable 
as predicted in many models of 
inflation~\cite{AAS79,RSV82,AW84,AGP85,CBDES93,CDS93}. 
These issues will be investigated in the future. 

\begin{acknowledgments}
This work was supported in part by JSPS Grants-in-Aid for 
Scientific Research 12640269 (M.S.) and 16340076 (J.Y.) 
and by Monbu-Kagakusho Grant-in-Aid for 
Scientific Research (S) 14102004 (M.S.). 
N.K. is supported by Research Fellowships of JSPS 
for Young Scientists (04249). 
\end{acknowledgments}

\clearpage

\clearpage

\clearpage

\clearpage

\clearpage

\clearpage

\clearpage

\clearpage

\clearpage

\end{document}